%% file: U15Dlayer.tex
\documentclass[a4paper,11pt,dvips]{article}
\usepackage{graphicx}
\usepackage{psfrag}
\usepackage{subfig}
\usepackage{amsmath,amstext,amsfonts,amsbsy,amssymb,amscd,bbm,epsfig}
\usepackage{verbatim}
\input{macros.tex}

\usepackage{cprform}
=-0.5cm \oddsidemargin=-0.5cm

\newcommand{\be}{\begin{equation}}
\newcommand{\ee}{\end{equation}}

\title{Establishment of the Coulomb law in the layer phase of a pure U(1) lattice gauge theory}
\author{K.~Farakos\footnote{E-mail: kfarakos@central.ntua.gr} and  
S.~Vrentzos\footnote{E-mail: vrentsps@central.ntua.gr}}

\begin{document}
\maketitle

\mbox{\normalsize Physics Department, National Technical University,15780 Zografou Campus, Athens, Greece} 


\begin{abstract}
In this article we examine the Layer phase of the five dimensional,
anisotropic, Abelian gauge model. Our results are to be compared with
the ones of the 4D U(1) gauge model in an attempt to verify  that  four 
dimensional physics  governs the four dimensional layers. The main results are
i) From the analysis of Wilson loops we verified the $\frac{1}{R}$ behavior,
in the layered phase, for the potential between heavy charges. The renormalized
 fine structure constant in the layer phase is found to be equal to that of 4D
Coulomb phase,$\alpha_{layer}$=$\alpha_{4D}$. ii) Based on the helicity modulus 
analysis we show that the layers are in the Coulomb phase while the transverse 
bulk space is in the confining phase. We also calculated the renormalized 
coupling $\beta_{R}$ and found results compatible with those obtained
from the Coulomb potential. Finally we calculated the potential in the 5D Coulomb 
phase and found $\frac{1}{R^{2}}$ behavior for the static $q \bar{q}$ potential.
From the study of the helicity modulus we have a possible estimate for 
the five dimensional renormalized fine structure constant in the region of the
critical value of the bare gauge coupling . 
\end{abstract}

\section{Introduction}
The idea that we live on a hypersurface embedded in a higher dimensional 
space
has attracted the interest of particle physicists and cosmologists in connection
with the hierarchy and the cosmological constant problems.
These ideas are motivated by string- and M-theory that are formulated in 
multidimensional
spaces. One version is to consider the extra dimensions flat, compactified to 
a large scale $M_{KK}\sim \frac{1}{R}$
varying from the Planck scale to a few TeV depending on the number of 
extra dimensions \cite{ADD,Dien}.
An alternative concept is the so called brane world scenario \cite{RSBO}, where all 
of the particles are localized on a
three-dimensional submanifold (brane), embedded in a multidimensional 
manifold (bulk), while gravitons are
free to propagate in the bulk. This model in five dimensions, implies a 
non-factorizable space
time geometry of the form
$AdS_{5}$ around the brane, assuming a negative five-dimensional cosmological 
constant .
The four-dimensional particles are expected to be gravitationally trapped 
on the brane.
Indeed gravitons and scalars for the second RS model \cite{RSBO} have a 
localized solution on the brane, plus a continuum spectrum .
Gravitons, scalars and fermions also exhibit a normalizable zero mode 
localized on the brane if we assume a non
trivial scalar background in the bulk, such as a kink topological defect 
toward the extra dimension \cite{Kehagias}.
The most difficult task is the massless non-abelian gauge field 
localization on the brane,
as any acceptable mechanism must preserve the charge universality 
\cite{Rubakov}.
A localization
mechanism that may be triggered by the extra dimensional gravity is 
proposed by the authors of refs \cite{FP,FP2,FP3}. It is based on the idea of the construction of a gauge 
field model which exhibits a non-confinement
phase on the brane and a confinement phase on the bulk, that is to say, a higgs mechanism 
driven by the coupling with the gravity.
Thus gauge fields, and more generally fermions and bosons with gauge 
charge, cannot escape into the bulk unless we
give them energy greater than the mass gap $\Lambda_{G}$, which emerges from the 
nonperturbative confining dynamics
of the gauge model in the bulk.
  In a previous work \cite{us} we have studied the 4+1 dimensional pure Abelian Gauge
model on the lattice with two anisotropic couplings, independent from each other and 
the coordinates, focusing our attention to the study of the phase diagram and the order
of the phase transitions. With this model we wanted to explore the possibility of a gauge
field localization scheme based on the observation that the anisotropy of the 
couplings produces a new phase, the layer phase, which mimics the  
Coulomb behavior in four dimensions but confines along the remaining one.
 This model is known since the mid-eighties when Fu and Nielsen proposed it as a new 
way to achieve dimensional reduction \cite{FuNi}. It is defined
on a D-dimensional space containing  d-dimensional subspaces. If the $\frac{d(d-1)}{2}$ 
couplings in the d-dimensional subspaces are identical ($\beta$) and the remaining $\frac{D(D-1)-d(d-1)}{2}$ coupling coefficients are also to be taken identical ($\beta^{'}$),
then for a certain range of parameters (typically $\beta^{'} \leq
\frac{1}{d}$ and $\beta \geq O(1)$)\footnote{This result comes from the mean field analysis of the theory.} this new phase emerges. The confinement along the $ (D-d) $ directions and the resulting detachment of the d-dimensional layers leads to the following physical picture. 
Charged particles in the layer phase will mainly run only along the layers since if they 
attempt to leave the layers in which they belong they will be driven back by a linear potential, analogous to the one responsible for quark confinement. Also gauge particles will follow the layers since there is no massless particle (photon) moving across the layers. 
We must note here however that
a stable layer phase exists only if $ D \geq 5 $ and $ d = D-1$ for the U(1) model.
For $ d \leq 3 $ we cannot have a layer phase since lattice gauge theory in less than four dimensions exhibits confinement for all finite values of the coupling constant, rendering the 
3-dimensional subspaces of the model incapable of  realizing a Coulomb phase . Also, due
to the asymmetric role of $\beta$ and $\beta^{'}$ in the  action, there is no layer phase
with their roles reversed.

 Many numerical investigations of the model have been made 
using Monte-Carlo techniques, verifying the structure of the phase diagram
and the properties of the different phases \cite{KoRAl,KKDF,FSDS}. In one of them \cite{KoRAl}
the presence of fermions in the model was investigated. The analysis, both analytical
(mean field approximation) and numerical (Monte Carlo simulations) revealed that the
qualitative characteristics of the phase diagram remained unchanged, even though a slight
restriction of the layer phase was observed in favor of the Coulomb one.
In \cite{KKDF} the authors analyzed the structure of  a U(1) model when the coupling $\beta^{'}$
in the fifth dimension depends on the coordinates exponentially, like in the RS model \cite{RSBO}.
In ref\cite{FSDS} the 5D anisotropic abelian Higgs model was analyzed and the existence
of a layer Higgs phase was established.  Finally, we would like to mention the main results in \cite{us} for the phase diagram of the pure U(1) anisotropic gauge model: i)
a weak first order phase transition between the 5D strong phase and the layer phase with the  characteristics of 4D U(1) and ii) strong indications for a second order transition between the layer phase and 5D Coulomb phase. In a preliminary study of the six dimensional U(1) model we
have found that the characteristics of the strong-layer phase transition remained the same
but at the same time the transition between the layer and the 6D Coulomb phase turned into
a strong first order one. 

For the case of the anisotropic non-Abelian SU(2) gauge model the
above picture changes \cite{RabBer}. Now the critical dimension for the formation of the
layers is D=6 giving as a minimal dimension for the layers d=5. But, as it is shown in ref \cite{DFKK}
a Layer Higgs phase in the non-abelian 5D model exists if one includes a scalar field in 
the adjoint representation.

In the present work we will focus our attention on the study of the long range
interactions of charged particles on the layers, at an attempt to further
justify our previous assumption \cite{us} that the layers incorporate all
the features that emerge in ordinary U(1) gauge theory in 4-dimensions. To this
end a very significant  step is the establishment of the Coulomb law and for that
we follow the usual approach: Measurements of Wilson loops (on the layers and 
for the 4-dimensional model), subsequent extraction of the potential and finally,
estimates for the string tension ($\sigma$) and the renormalized charge or fine
structure constant ($\alpha$) are obtained. 
An equally important byproduct of the above analysis is determination of the role
of layer-layer interactions and their consequences (if any) for the physical picture
in the layers.

  Our work is organized as follows. In section 2 we present the four dimensional
model along with the various analysis techniques used throughout this paper. In 
section 3 we present the layer phase results and a comparison with their 4D 
counterparts and finally, in section 4, we concentrate  in the 5D Coulomb phase 
in order to show the clear distinction (both qualitative and quantitative) from the
layer one as well as an attempt to explore the very nature of what should be a 
five-dimensional Coulomb law.

\section{The 4-dimensional case}
\subsection{Wilson loops and the static potential}
  One of the observables used in lattice gauge theories with great physical
significance, is the Wilson loop defined as the gauge invariant quantity consisting of an
ordered product of link variables along a contour C. 
If we denote 
by $U_{l}$ such a link variable then the Wilson operator is defined as 
\begin{equation}
W_{C} = \prod_{l  \in C} U_{l}
\end{equation}
and its expectation value on a rectangular loop C is: 
\begin{equation}
 W(\widehat{R},\widehat{T}) \equiv \langle W_{C}[U] \rangle,
\end{equation}
where $\widehat{R}$ and $\widehat{T}$ are the (dimensionless) spatial and
temporal extention of the  contour C. The symbol $\left < \dots \right > $ 
denotes the expectation value with respect to the 4d gauge action:\\
\begin{displaymath}
 S^{4D}=\beta \sum_{x, \mu \leq \nu}(1-Re(U_{\mu \nu}(x)))
\end{displaymath}\\

From the asymptotic behavior of the above quantity we can  (in principle) derive 
the potential between two static charges using numerical methods through the
formula
\begin{equation}
 \widehat{V}(\widehat{R})= - \lim\limits_{\widehat{T}\to \infty}\frac{\ln W(\widehat{R},\widehat{T})}{\widehat{T}}
\end{equation}

There are also R-independent self energy contributions to $\widehat{V}(\widehat{R})$
that one has to take into account (see next section).

\subsection{Helicity Modulus and the renormalized coupling}
A very usefull quantity for the characterization of phases in lattice gauge theories
is the helicity modulus ($\bf{h.m}$), first introduced in this context 
by P.de Forcrand and M.Vettorazzo \cite{PFMV}. It characterizes the responce of a 
system to an external flux and has the behavior of an order parameter.
It is zero in a confining phase and nonzero in a coulombic one \cite{PFMV,Cardy}.\\
The helicity modulus is defined as:
\begin{equation}
h(\beta)= \left. \frac{\partial^{2}F(\Phi)}{\partial \Phi^{2}}\right| _{\Phi=0}
\end{equation}
were $\Phi$ is the external flux and $F(\Phi)$ the flux free energy given by:
\begin{equation}
 F(\Phi)=-\ln\left(Z(\Phi)\right),\quad Z(\Phi)= \int D\theta  \mbox{\large e}^{\sum_{stack}(\beta\cos(\theta_{P}+\Phi)) + \sum_{\overline{stack}}
(\beta\cos(\theta_{P}))}
\end{equation}
with $Z(\Phi)$ the partition function of the system due to the presence of the external flux.
\begin{math} 
\sum_{stack}
\end{math}
is the sum over the stack of plaquettes, of a given orientation (e.g $\mu,\nu$)
in which the extra flux is imposed ($\theta_{P} \rightarrow \theta_{P}+\Phi$), and 
\begin{math}
 \sum_{\overline{stack}}
\end{math}
is its complement, consisting of all the plaquettes that remained unchanged.\\
An observation that will subsequently play an important role is that the partition
function $Z(\Phi)$ of equation (2.5) and hence the flux free energy is clearly 
$2\pi$ periodic . So, the extra flux we impose on the system is defined only
 $mod(2\pi)$.

If we, take for example: 
\begin{math}
 \mbox{stack}=\{\theta_{\mu \nu}(x,y,z,t) | \mu=1, \nu=2;x=1, y=1\}
\end{math}
 then, with a suitable change in variables we can spread the extra flux uniformly to 
all the plaquettes in the ($\mu-\nu$) plane. The partition function now becomes
\begin{equation}
 Z(\Phi)= \int D\theta \mbox{\large e}^{\beta\sum_{(\mu \nu)planes} \cos(\theta_{P}+\frac{\Phi}{L_{\mu}L_{\nu}})+\beta\sum_{(\overline{\mu \nu})planes}\cos(\theta_{P})}
\end{equation}
and from equation (2.4) we get for the h.m:
\begin{equation}
 h(\beta)=\frac{1}{(L_{\mu}L_{\nu})^{2}}\left( \left< \sum_{(\mu \nu)\mbox{\tiny planes}}(\beta\cos(\theta_{P}))\right>
-\left< (\sum_{(\mu \nu)\mbox{\tiny planes}}(\beta\sin(\theta_{P})))^{2} \right>\right)
\end{equation}
with the sum extending to all planes parallel to the given orientation.

Now, consider for the moment the classical limit ($\beta\rightarrow\infty$) for the action
of equation (2.6) where all the fluctuations are suppressed. In
this limit the flux is distributed equally over all the plaquettes of each plane and
does not change as we cross parallel planes. If we expand the classical action in powers
of the flux \cite{PFMV}, since in the thermodynamic limit $\frac{\Phi}{L_{\mu}L_{\nu}}$ 
is always a small quantity, we find:
\begin{displaymath}
 S_{\mbox{\tiny classical}}(\Phi)=\frac{1}{2}\beta \Phi^{2}\frac{V}{(L_{\mu}L_{\nu})^{2}}\mbox{\normalsize + constant}\Longrightarrow
F_{\mbox{\tiny classical}}(\Phi)-F_{\mbox{\tiny classical}}(0)=\frac{1}{2}\beta \Phi^{2}\frac{V}{(L_{\mu}L_{\nu})^{2}}
\end{displaymath}
where V is the lattice volume, $V=L_{\mu}L_{\nu}L_{\rho}L_{\sigma}$.

The above expression for the free energy F holds 
 all the way up to the phase transition, where fluctuations are present, if one only replaces
the bare coupling by a renormalized coupling: $\beta \rightarrow \beta_{R}(\beta)$
(for details see \cite{PFMV,Cardy}).

Upon replacing $\beta_{R}(\equiv\frac{1}{e_{R}^{2}})$ $\rightarrow$ $\frac{1}{4\pi\alpha_{R}}$ the above expression becomes:
\begin{equation}
F_{[\mbox{\tiny finite $\beta$}]}(\Phi)-F_{[\mbox{\tiny finite $\beta$}]}(0)=
\frac{\beta_{R}}{2}\Phi^{2}\left(\frac{L_{\rho}L_{\sigma}}{L_{\mu}L_{\nu}}\right)=
\frac{\Phi^{2}}{8\pi\alpha_{R}}\left(\frac{L_{\rho}L_{\sigma}}{L_{\mu}L_{\nu}}\right)
\end{equation}
The above equation does not show any periodicity in $\Phi$. To remedy this situation
we have to consider all configurations whose flux is a multiple (k) of $2\pi$.
\begin{equation}
F(\Phi)=-\mbox{ln}\left(\sum_{k}\mbox{\large e}^{-\frac{\beta_{R}}{2}\left(\frac{L_{\rho}L_{\sigma}}{L_{\mu}L_{\nu}}\right) (\Phi+2 \pi k)^{2}}\right)
\end{equation}
Now we can define $\beta_{R}(\beta)$ implicitly from the equation:
\begin{equation}
 \left.\frac{\partial^{2}F(\Phi,\beta_{R})}{\partial \Phi^{2}}\right| _{\Phi=0}=h_{0}(\beta)
\equiv h(\beta)\quad \mbox{or alternatively, }  \left.\frac{\partial^{2}F(\Phi,\beta_{R})}{\partial \Phi^{2}}\right| _{\Phi=\pi}=h_{\pi}(\beta)
\end{equation}
As equations (2.9) and (2.10) show,  the renormalized coupling equals
the helicity modulus up to exponentially small corrections \cite{PFMV}.

\subsection{Measurements}
  Our Monte Carlo calculations for the case of four-dimensional QED are 
restricted to volumes $V=12^{4}$, $14^{4}$ and $16^{4}$. For all lattice sizes,
the work of Jersak et al \cite{Jersak} has been closely followed. We used a 5-hit
Metropolis algorithm supplemented by an overrelaxation method. About $10^{5}$
sweeps were used for thermalization and more than 2 x $10^{5}$ measurements,
10 sweeps apart from each other, for the determination of mean values.
For the case of $V=16^{4}$ all planar rectangular Wilson loops with R=1,...,6
and T=1,...8 were calculated while for the rest we used 
$R,T < \frac{L}{2}$ loops in an effort to minimize  finite size effects.

In general, one has to extract the potential from the logarithms of the 
expectation values of Wilson loops for large T.

\begin{equation}
 -\ln\langle W_{C}[U] \rangle = V(R)\times T +const
\end{equation}

This however requires a large enough volume to deal with the many issues that
emerge in lattice calculations. Finite size effects are a constant ``threat''
to the validity of the results  
and in addition finite T effects  must also be taken into account in the 
extraction of the potential. This means that significant deviations from a 
linear dependence on T should  be investigated.
To this end, a third  term in the above equation is introduced

\begin{equation}
 -\ln\langle W_{C}[U] \rangle = const + V(R)\times T + \frac{C}{T}
\end{equation}
with C actually being a function of R.

The form of the ``correction'' term is an open question but we choose  
$\sim\frac{1}{T}$ as it is the simplest (and most obvious) choice. The 
potential V(R) has been calculated using both linear (eq. (2.11)) and
non-linear (eq. (2.12)) dependence of the logarithms on T, at a variable 
number of points dependent on the volume under consideration, giving
comparable results within errors. The resulting
values were fitted to a superposition of  linear + Coulomb potentials :

\begin{equation}
 V(R)=\sigma_{cc} R - \frac{\alpha_{cc}}{R} + const 
\end{equation}
and 
\begin{equation}
V(R)=\sigma_{lc} R -\alpha_{lc} V_{lc}(R) +const
\end{equation}
with $V_{lc}(R)$  the lattice Fourier transform of a massless bosonic propagator \cite{Jersak,CHMV}
which respects not only the momentum cut-off but also accounts for the periodicity of the 
lattice. 
\begin{equation}
 V_{lc}(R)=\frac{4 \pi}{L_{s}^{3}}\sum_{\overrightarrow{k}\neq 0}\frac{\mbox{\large e}^{i\vec{k}\vec{R}}}{\sum_{j=1}^{3}2(1-\cos(k_{j}))}, \qquad k_{j}=0,\frac{2\pi}{L_{s}},\dots,\frac{2\pi(L_{s}-1)}{L_{s}}
\end{equation}

All measurements focused on the Coulomb phase starting from values near 
the critical point and extending to larger values of $\beta$.
In Table 1 we present the results for several volumes  using both the continuum (eq. (2.13))
and the lattice Coulomb potential (eq. (2.14)) obtained from the non-linear fitting (eq. (2.12)).
The notation we use is indicative. With $\bf{cc}$ we imply the use of continuum Coulomb
potential ($\frac{1}{R}$) and with $\bf{lc}$ the use of lattice Coulomb potential (eq. (2.15)).
A few remarks are now in order.
First of all, the estimates of $\alpha$ using
the two different types of potential, show a systematic deviation of the order 
of 0.010 - 0.020 which was also found to be true by Jersak et al \cite{Jersak}. 
Second, it is evident from Table 1 that the values of $\alpha$ show a quick convergence
to the infinite volume limit as their deference, between the biggest volumes
that we used ($14^{4}$ and $16^{4}$) is well within errors. The string tension starts
at relatively large values due to finite size effects for the smaller system under 
study, only to reach a final value of 0.003 as the volume increases. This result
is  slightly bigger than the one found by Jersak et al. It appears that
the reason for this systematic overestimation of $\sigma$ has its origin at
the insertion of the extra term ($\sim\frac{1}{T}$) of equation (2.12). This extra term 
inserts an ''effective'' string tension that adds up to the measurement, but this was anticipated. In ref \cite{CHMV} it was found that the static charge potential 
obtained from Wilson loops, acquires a confining contribution $\simeq \frac{cR}{T^{2}}$ 
for finite volume. This term has exactly the same form as the term ($\frac{C}{T}$) that
we have added by hand and
consequently  we have enhanced an already present confining contribution to the potential.
One could in general monitor the contribution of this extra term and subtract it  in order to remedy the presented inconsistency between the values of the string tension ($\sigma$) as they are obtained through the use of equations (2.11) and (2.12) (Tables 1 \& 2). Unfortunately, knowledge of the precise functional dependence of C from R is required, a topic that proves to be not an easy task. So, in order to extend our measurements for the renormalized charge to smaller 
volumes we have enhanced the signal for the string tension. This however does not affect
the obtained values of $\alpha$, a fact most apparent in the subsequent analysis.
In Table 2 we present our results for a linear fit (equation (2.11))
and L=16 for both types of potential ($\bf{c}$ontinuum $\bf{C}$oulomb and $\bf{l}$attice
$\bf{C}$oulomb). Comparing with Table 1, for L=16, one can see that by reaching
a big enough volume the extra term does not play any substantial role. Insofar as 
$\alpha$ is concerned, linear and non-linear fits give exactly the same results 
(within errors) (Figure.1). The string tension $\sigma$ is found smaller and 
compatible with zero within the statistical errors. The linear fit gives us a more
convincing signal for the vanishing of the string tension with values comparable 
with the ones found by Jersak et al \cite{Jersak}. 
\begin{table}[ht]
 \caption{Results for the Non-linear fitting}

\begin{center}
\begin{tabular}{|l|c|c|c|r|}
\hline
 \multicolumn{5}{|c|}{L=16} \\
\hline
$\beta$ & $\alpha_{cc}$ &$\sigma_{cc}$ &$\alpha_{lc}$ &$\sigma_{lc}$ \\
\hline   
1.015 & 0.1860(20) & 0.0033(5) & 0.1693( 9) & 0.0027(3) \\
\hline
1.020 & 0.1765(16) & 0.0035(4) & 0.1605( 9) & 0.0030(3) \\
\hline
1.030 & 0.1631(13) & 0.0036(3) & 0.1484(11) & 0.0030(3) \\
\hline
1.050 & 0.1472(12) & 0.0033(3) & 0.1339(11) & 0.0027(3) \\
\hline
 \multicolumn{5}{|c|}{L=14}   \\
\hline
$\beta$ & $\alpha_{cc}$ &$\sigma_{cc}$ &$\alpha_{lc}$ &$\sigma_{lc}$ \\
\hline
1.015   & 0.1883(34) & 0.0039(10)  & 0.1692(12) & 0.0024(3)\\
\hline
1.025   & 0.1716(26) & 0.0040(10)  & 0.1540(11)& 0.0027(3)\\
\hline
1.030   & 0.1632(23) & 0.0034(10)  & 0.1474(14)& 0.0023(4) \\
\hline
1.040   & 0.1577(29) & 0.0035(10)  & 0.1414(12)& 0.0027(3)\\
\hline
1.050   & 0.1472(21) & 0.0045(15)  & 0.1329(14)& 0.0020(4) \\
\hline
 \multicolumn{5}{|c|}{L=12}   \\
\hline
$\beta$ & $\alpha_{cc}$ &$\sigma_{cc}$ &$\alpha_{lc}$ &$\sigma_{lc}$ \\
\hline
1.013   & 0.1975(2) & 0.0060(10) & 0.1765(12) & 0.0041(3)\\
\hline
1.015   & 0.1906(2) & 0.0060(10) & 0.1708(14) & 0.0040(3)\\
\hline
1.025   & 0.1730(2) & 0.0055(20) & 0.1552(12) & 0.0040(2)\\
\hline
1.050   & 0.1537(2) & 0.0055(10) & 0.1366(10) & 0.0038(2)\\
\hline
1.060   & 0.1466(2) & 0.0055(10) & 0.1307(10) & 0.0034(2)\\
\hline
1.100   & 0.1338(2) & 0.0050(10) & 0.1182(10) & 0.0035(2)\\
\hline 
1.200   & 0.1087(2) & 0.0050(10) & 0.0968(10) & 0.0026(2)\\
\hline
 \end{tabular}
\end{center}
\end{table}

\begin{table}[h!]
\caption{Results from the Linear Fit and L=16}
\begin{center}
\begin{tabular}{|l|l|c|r|r|}
\hline
\multicolumn{5}{|c|}{L=16} \\
\hline
$\beta$ & $\alpha_{cc}$ & $\sigma_{cc}$ & $\alpha_{lc}$ & $\sigma_{lc}$ \\
\hline   
1.015 & 0.1868(30) & 0.0010(7) & 0.1698(23) & 0.0020(6) \\
\hline
1.020 & 0.1771(26) & 0.0005(6) & 0.1610(23) & 0.0010(6) \\
\hline
1.030 & 0.1638(22) & 0.0003(5) & 0.1488(20) & 0.0010(5) \\
\hline
1.050 & 0.1476(20) & 0.0003(4) & 0.1342(18) & 0.0008(5) \\
\hline
 \end{tabular}
\end{center}
\end{table}
\newpage
\begin{figure}[ht!]
\begin{center}
\includegraphics[width=9cm,angle=270]{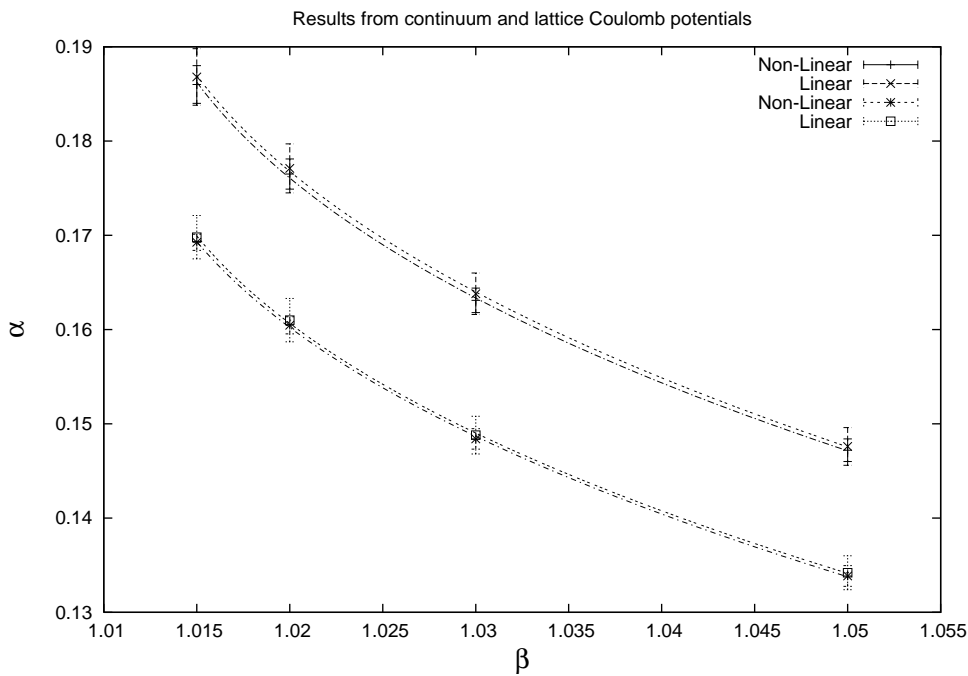}
\caption[]{Results from linear and non-linear fits from the $16^{4}$ lattice for 
the continuum Coulomb (upper curves) and lattice Coulomb (lower curves) potentials respectively}
\end{center}
\label{f-2-1}
\end{figure}
In 1982 Luck \cite{Lck} using the close analogy between the 2-dimensional XY
model and 4-dimensional compact QED arrived by means of a weak-coupling 
expansion, at the following form for the behavior of the renormalized fine 
structure constant
\begin{equation}
 \alpha(\beta)=\alpha_{c} -const\times(1-\frac{\beta_{c}}{\beta})^{\lambda}
\end{equation}
with $\alpha_{c}\simeq 0.15$ and $\lambda\simeq 0.5$.
The analysis lead him to the conclusion that the square of the renormalized charge
takes a universal value $e_{c}^{2} \equiv 4 \pi \alpha_{c}$=1.90$\pm$ 0.10 at the 
deconfinement point.

By making use of equation (2.16) and the most accurate (to our knowledge) value for the
critical  $\beta$ in four dimensions ($\beta_{c}=1.0111331(21)$ \cite{ABLS}) we present
in Table 3 our results for $\alpha_{c}$ and $\lambda$, were the last row (L=16) refers 
to the results of the linear fit and the  
L=10 entry actually amounts to a $16\times10^{3}$ lattice volume. Our values are in very
good agreement with those found in ref\cite{Jersak} and those predicted through 
theoretical calculations \cite{Lck}. The systematic error in the analysis due to the 
presence of $\delta \beta_{c}$ turns out to be insignificant. 

\begin{table}[h!]
\caption{Results for $\alpha_{c}$ and $\lambda$}
\begin{center}
 \begin{tabular}{|l|l|c|c|r|r|}
 \hline
L  & $\alpha_{c-cc}$ & $\alpha_{c-lc}$ & $\lambda_{cc}$ & $\lambda_{lc}$ & $\chi^{2}_{d.o.f}$ \\
\hline
10 & 0.230(25) & 0.208(24)         & 0.31( 8)  & 0.33(10) & 1.05        \\
\hline 
12 & 0.209( 7)      & 0.200( 4)      &  0.38( 5) & 0.34( 3) & 0.80          \\
\hline
14 & 0.211(23)     & 0.190(10)     &  0.44(20) & 0.45(13) & 0.75         \\
\hline
16 & 0.211(16)     & 0.190( 9)     & 0.42(16)  & 0.43(10)& 0.20          \\
\hline
16 & 0.211(23)     & 0.192(21)    & 0.43(10)  & 0.42(15)& 0.10          \\
\hline
\end{tabular}
\end{center}
\end{table}

 A pleasing fact is
that both types of potential manage to describe equally well our data, giving 
identical results (within errors) and thus providing us with a signal for the
existence of a massless photon in the Coulomb phase. The only noticeable difference
between the two sets  comes from the appearance of a systematic volume dependence for 
$\alpha_{c-lc}$, with evidence that better accuracy is provided by the lattice potential,
as it better takes into account the system volume. Finally, the inclusion of the extra term
($\sim \frac{1}{T}$) proved quite efficient, since it allowed us to obtain the required information even at smaller volumes. 

\section{The 5-dimensional anisotropic model, layer phase}
\subsection{The model}
In this section we consider the five dimensional anisotropic U(1) lattice gauge model
with two couplings $\beta$ and $\beta^{'}$: 
\begin{equation}
S^{5D}_{gauge}=\beta \sum_{x,1 \leq \mu < \nu \leq 4}(1- Re(U_{\mu\nu}(x))+
\beta^{'} \sum_{x,1 \leq \mu \leq 4}(1- Re(U_{\mu 5}(x))
\end{equation}
where 
\begin{displaymath}
\begin{array}{ccc}
U_{\mu\nu}(x)& = & U_{\mu}(x)U_{\nu}(x+\alpha_{s}\hat{\mu})U_{\mu}^{\dagger}(x+\alpha_{s}\hat{\nu})U_{\nu}^{\dagger}(x)\\
U_{\mu 5}(x) & = &U_{\mu}(x)U_{5}(x+\alpha_{s}\hat{\mu})U_{\mu}^{\dagger}(x+\alpha_{5}\hat{5})U_{5}^{\dagger}(x)
\end{array}
\end{displaymath}
\\
are the plaquettes defined on the 4-d subspace ($\mu,\nu$ = 1,2,3,4) and on the plane 
containing the transverse fifth direction ($x_{5}$) respectively .
\footnote{By $\alpha_{s}$ and $\alpha_{5}$ we denote the two different lattice spacings:
one referring to the 4-dimensional subspaces and the other to the transverse fifth direction.}
The link variables are defined as:
\begin{displaymath}
 U_{\mu}=\exp(i\theta_{\mu}(x)),\quad U_{5}=\exp(i\theta_{5}(x))
\end{displaymath}
and in terms of them the plaquette variables can be written as
\begin{displaymath}
 U_{\mu\nu}(x)=\exp(i\theta_{\mu\nu}(x)),\quad U_{\mu 5}(x)=\exp(i\theta_{\mu 5}(x))
\end{displaymath}
with the definitions
\begin{displaymath}
\begin{array}{ccc}
 \theta_{\mu\nu} & = & \theta_{\mu}(x)+\theta_{\nu}(x+\alpha_{s}\hat{\mu})-\theta_{\mu}(x+\alpha_{s}\hat{\nu})
-\theta_{\nu}(x) {}\\
\theta_{\mu 5} & = & \theta_{\mu}(x) + \theta_{5}(x+\alpha_{s}\hat{\mu})-\theta_{\mu}(x+\alpha_{5}\hat{5})
-\theta_{5}(x)
\end{array}
\end{displaymath}
Before we proceed we would like to define the helicity modulus for this model.
The anisotropy of the couplings and the resulting enrichment of the phase 
diagram introduces the necessity for two kinds of h.m. One probing the response
of the system to an external flux through the spatial planes ($\mu-\nu$) and
one for the transverse planes ($\mu-5$).   
\begin{equation}
 h_{s}(\beta)=\frac{1}{(L_{\mu}L_{\nu})^{2}}\left(\left<\sum_{P}(\beta\cos(\theta_{\mu\nu}))
\right>-\left<(\sum_{P}(\beta\sin(\theta_{\mu\nu}))^{2})\right>\right)
\end {equation}
\begin{equation}
h_{5}(\beta^{'})=\frac{1}{(L_{\mu}L_{5})^{2}}\left(\left<\sum_{P^{'}}(\beta^{'}\cos(\theta_{\mu 5}))
\right>-\left<(\sum_{P^{'}}(\beta^{'}\sin(\theta_{\mu 5}))^{2})\right>\right)
\end{equation}
with the sum of equation (3.3) extending on all the plaquettes on the transverse plane.

  The phase diagram of this model includes three distinct phases (Fig.2). For large
values of the couplings ($\beta$, $\beta^{'}$) the model lies in a $\bf{C}$oulomb phase 
on a 5D space. Now, with $\beta$ fixed, as $\beta^{'}$ decreases the system will eventually develop a behavior according to which the force in four dimensions will be Coulomb-like 
while in the fifth direction the system will exhibit confinement. This is the new $\bf{L}$ayer
phase. For small values of both $\beta$ and $\beta^{'}$ the force is confining in all
five directions and the coresponding phase is the $\bf{S}$trong phase.
  \begin{figure}[h!]
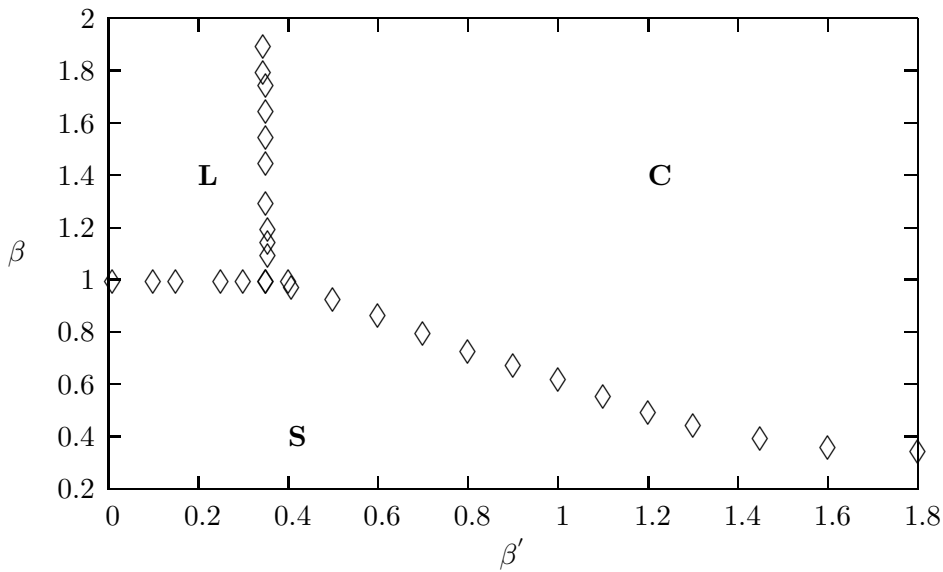

\begin{center}
 \include{PDG}
\caption[]{The phase diagram of the theory}
\label{f-3-1}
\end{center}
\end{figure}

The Wilson loops and the helicity moduli are expected to exhibit different behavior as one
crosses the phase boundaries. In the $\bf{S}$trong (confinement) phase all Wilson loops obey
the area law while at the same time the helicity modulus is zero throughout the appropriate range of parameters, both signals for confinement. In the 5D $\bf{C}$oulomb phase  the   opposite picture emerges. Wilson loops obey the perimeter law with the helicity modulus
being nonzero and scaling with the lattice length as $\beta$ and $\beta^{'}$ increase. A five
dimensional Coulomb-type force is present. Finally, the $\bf{L}$ayer phase consists of a
mixture of both aforementioned phases. The Wilson loops constrained in the 4d subspaces
($W_{\mu\nu}$ with $1\leq\mu,\nu\leq 4$) obey the perimeter law while at the same time those that contain the fifth direction ($W_{\mu 5}$, $1\leq\mu\leq 4$) show an area
law behavior. The helicity modulus shows also two different behaviors. The space h.m ($h_{S}(\beta$)) has a nonzero value in the layer phase while the transverse h.m ($h_{5}(\beta^{'})$) is constrained to a zero value as one would expect from a confining 
force (Figure 8).  

\subsection{Measurements}
The calculations of this section are dedicated entirely to the layer phase for
the  range of parameters $\beta^{'}$=0.2 and $1.015\leq \beta \leq 1.40$.
In order to illustrate the qualitative and quantitative agreement between the
layer phase of the 5 dimensional model and the corresponding 4d systems we 
focused  on  $16 \times 10^{4}$ and $12^5$ volumes, which in the context of the
layer phase translates to ten and twelve layers of volume $16\times10^{3}$ and $12^4$ 
each\footnote{In the notation that will be used from now on 4D will signify the four
dimensional model while 4d the four dimensional subspaces (layers) of the five 
dimensional system.}
. Every layer is
(to a very large extent) decorrelated \cite{FSDS} from the others and every quantity measured
on it a random variable with a given distribution. So, it really does not matter
which layer we choose to observe, since each one of them will demonstrate exactly the
same behavior. If we treat the  system as a whole this would only amount to an
increase in statistics. In order to probe the
physics in the layers  all planar rectangular Wilson loops with R=1,...,5 and 
T=1,...,8 and R=1,...,6 and T=1,...,6 depending on the case under study
were constructed from link variables living only on the 4-dimensional
subspaces  ($U_{\mu},\quad \mu=1,2,3,4$)
\footnote{Although finite size effects forced us to disregard all borderline sizes.}
while at the same time independent runs were made to the corresponding  4-dimensional models
($V=16\times10^{3}$, V=$12^4$) for a straightforward comparison. Following the same steps
 as in the previous section we investigated the long range correlations, in terms
of the dimensionless parameter $\alpha(\beta)$, in these two different systems. 
\subsubsection{\mbox{The 16 $\times$ $10^4$ case}}
We use equation (2.12) in order to extract 
the potential from the mean values of the Wilson loops.
All points with T=1,2 were excluded from the fits and even T=3 for $R\geq4$ (Fig.3)
\cite{Jersak}. 
We were able to determine the potential V(R) only at 4 points R=1,...,4 for each
value of $\beta$ (because of the ``noise'' introduced by finite size effects) and
compare these values with the ones  from the 4-dimensional model. 
\begin{figure}[!hts]
\begin{center}
\includegraphics[width=7cm,angle=270]{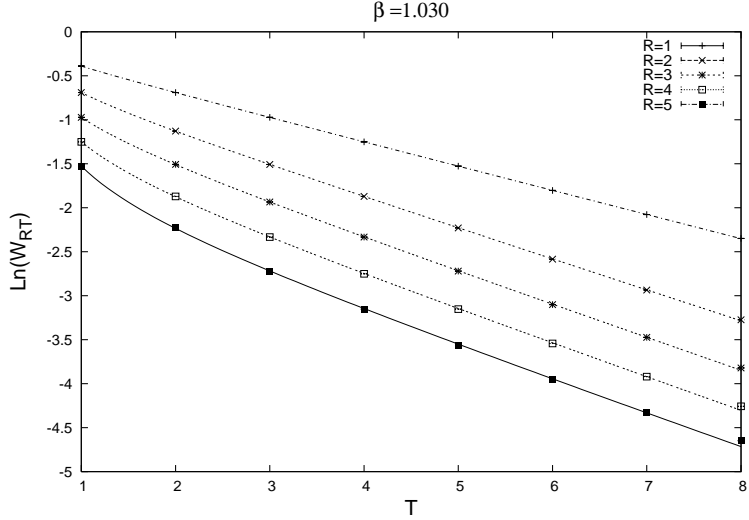}
\caption[]{The logarithms of the expectation values for Wilson loops at $\beta$=1.030 and
lattice volume V=$16\times 10^{4}$.The lines are the result of the fitting with equation (2.12).
The error bars are included in the symbols size.}
\end{center}
\label{f-3-2}
\end{figure}

\begin{figure}[!h]
\centering
\subfloat[The potentials in the layer phase,$V_{5D}=16\times 10^{4}$ from $\beta$=1.015
(upper curve) to $\beta$=1.070 (lower curve). The lines represent
a fit with equation (2.13) using the continuum Coulomb form]
{\includegraphics[width=5cm,angle=270]{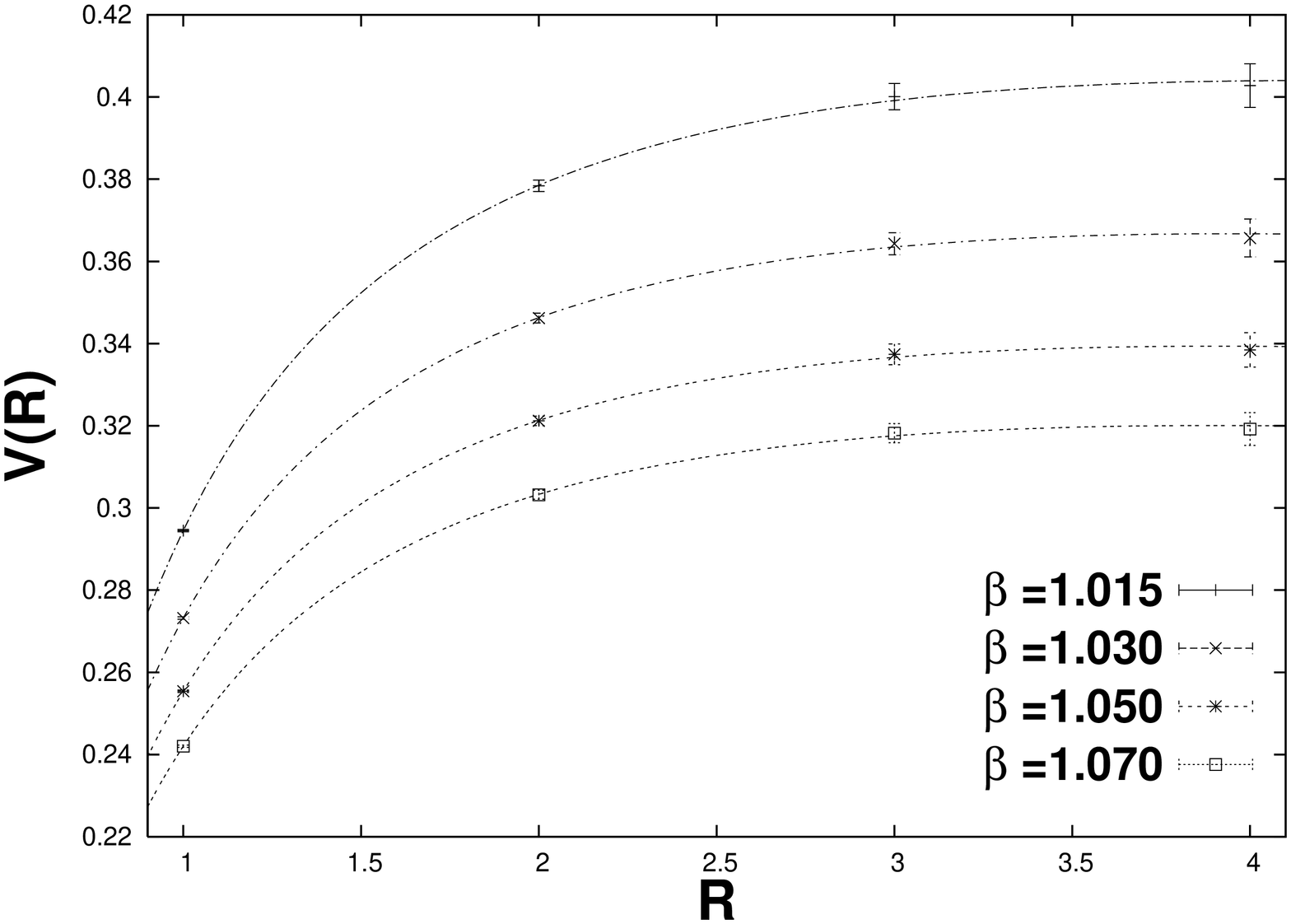}}\quad
\subfloat[Comparison of the 4d layer potential with the usual 4D potential
for $\beta$=1.015, as obtained from equation (2.13) using the continuum Coulomb form.]
{\includegraphics[width=5cm,angle=270]{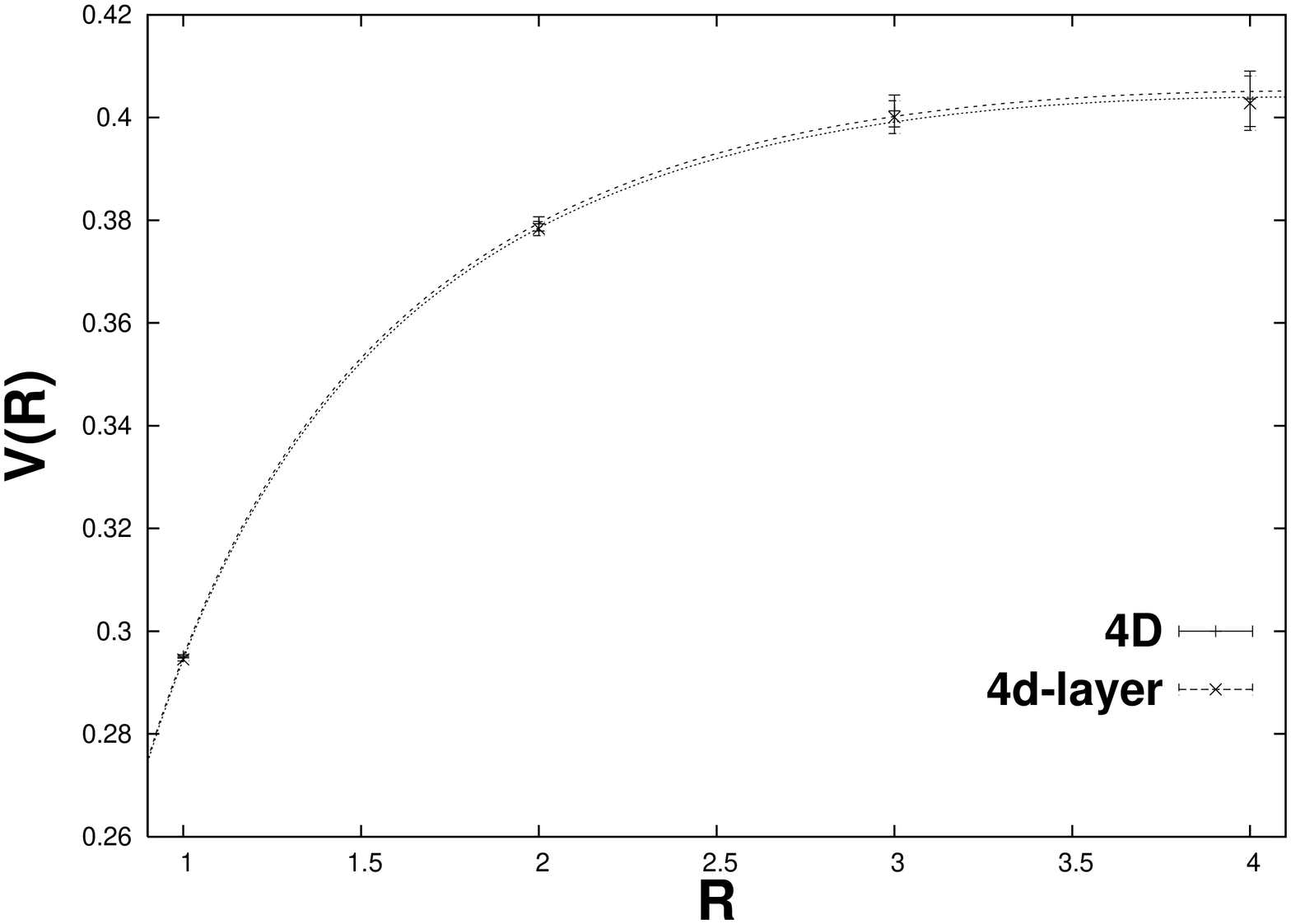}}
\caption{}
\end{figure}
The obtained values were fitted to a superposition of a linear + $\bf{C}$oulomb potentials
for both forms of the latter ($\bf{c}$ontinuum and $\bf{l}$attice). The $\sim \frac{1}{R}$
behavior  describes well the data (Fig.4(a)) and gives results 
compatible with the ones obtained from the 4D model (Fig.4(b)). This serves as a first 
signal for the presence of a four dimensional Coulomb law in the layer phase. The second,
and most important, is the fact that equally good results are provided by means of the four 
dimensional lattice propagator (eq. (2.15)), a quantity that describes the long range 
interactions in four dimensional lattices. The success in the description of the data comes
as strong evidence of the four dimensional nature of the layers.

So, it seems that both signals, the $(\frac{1}{R})$ form of the potential in the layers
and the success of the massless bosonic propagator in the description of the data, which
by itself could be considered as evidence of the presence of a massless boson acting as
mediator to the forces in the layer, point to the existence of a  4d gauge particle in 
the layer phase with all the characteristics of an ``ordinary'' photon.
In Tables 4 and 5 we present the results for $\alpha$ and $\sigma$ for the two models,
$16\times 10^{4}$ and $16\times 10^{3}$.
The similarity between the two sets of measurements is very encouraging. The two systems
reveal exactly the same behavior (as it is demonstrated by the two measured quantities)
regardless of the form chosen for the Coulomb potential.  



\begin{table}[ht]
 \caption{Results from the layer phase using the continuum Coulomb potential}
\begin{center}
\begin{tabular}{|l|c|c|c|r|}
\hline
   & \multicolumn{2}{l|}{\qquad $V=16\times10^{4}$}      & \multicolumn{2}{c|}{$V=16\times10^{3}$}       \\
\hline
$\beta$ & $\alpha_{\mbox{\tiny layer}}$ &$\sigma_{\mbox{\tiny layer}}$ &$\alpha_{4D}$ &$\sigma_{4D}$ \\
\hline   
1.015   & 0.1910(88) & 0.0110(35) & 0.1907(80) & 0.0110(33)\\
\hline
1.030   & 0.1677(78) & 0.0107(30) & 0.1683(75)      & 0.0108(30)\\
\hline
1.050   & 0.1517(55) & 0.0098(32) & 0.1522(68) & 0.0100(27)\\
\hline
1.070   & 0.1411(64) & 0.0093(26) & 0.1412(67) & 0.0093(26)\\
\hline
1.080   & 0.1367(62) & 0.0090(25) & 0.1370(63) & 0.0090(25)\\
\hline
1.090   & 0.1330(62) & 0.0087(24) & 0.1332(20) & 0.0088(25)\\
\hline
1.100   & 0.1295(63) & 0.0085(23) & 0.1298(30) & 0.0083(19) \\
\hline
1.200   & 0.1101(40) & 0.0079(23) & 0.1093(23)  & 0.0070(20)\\
\hline 
1.300   & 0.0900(100) & 0.0065(35) & 0.0932(17) & 0.0060(18)\\
\hline
1.400   & 0.0830(40) & 0.0054(22) & 0.0822(40) & 0.0049(16)\\
\hline
 \end{tabular}
\end{center}
\end{table}

 \begin{table}[ht]
 \caption{Results using the lattice Coulomb potential}
\begin{center}
\begin{tabular}{|l|c|c|c|r|}
\hline
   & \multicolumn{2}{c|}{$V=16\times10^{4}$}     & \multicolumn{2}{c|}{$V=16\times10^{3}$}       \\
\hline
$\beta$ & $\alpha_{\mbox{\tiny layer}}$ &$\sigma_{\mbox{\tiny layer}}$ &$\alpha_{4D}$ &$\sigma_{4D}$ \\
\hline   
1.015   & 0.1747(80) & 0.0097(34) & 0.1753(73) & 0.0097(34)\\
\hline
1.030   & 0.1541(70) & 0.0096(29) & 0.1546(70) & 0.0096(28) \\
\hline
1.050   & 0.1394(64) & 0.0089(27) & 0.1397(63) & 0.0090(27)\\
\hline
1.070   & 0.1296(61) & 0.0083(25) & 0.1322(69) & 0.0083(25)\\
\hline
1.080   & 0.1256(59) & 0.0080(24) & 0.1258(59) & 0.0081(25)\\
\hline
1.090   & 0.1221(59) & 0.0078(24) & 0.1223(54) & 0.0079(24)\\
\hline
1.100   & 0.1189(56) & 0.0075(23) & 0.1192(50) & 0.0075(22) \\
\hline
1.200   & 0.1065(60) & 0.0068(24) & 0.0981(48) & 0.0062(20)\\
\hline 
1.300   & 0.0792(96) & 0.0057(20) & 0.0855(44) & 0.0054(17)\\
\hline
1.400   & 0.0763(38) & 0.0048(16) & 0.0754(39) & 0.0043(16)\\
\hline
 \end{tabular}
\end{center}
\end{table}
\newpage 
We repeat the whole analysis for the linear case using equation (2.11)
since our ``correction'' term only affects the T direction and, as is evident from our
4-dimensional study, the value T=16 proves to be sufficient. The agreement between
the relevant sets of measurements (Tables 6 \& 7) is extremely good (indistinguishable 
within the errors) as one can also see in Figures 5(a) \& 5(b). 

 \begin{table}[ht!]
 \caption{Results for the continuum Coulomb potential, linear fits}
\begin{center}
\begin{tabular}{|l|c|c|c|r|}
\hline
    & \multicolumn{2}{c|}{$V=16\times10^{4}$}      & \multicolumn{2}{c|}{$V=16\times10^{3}$} \\
\hline
$\beta$ & $\alpha_{\mbox{\tiny layer}}$ &$\sigma_{\mbox{\tiny layer}}$ &$\alpha_{4D}$ &$\sigma_{4D}$ \\
\hline   
1.015   & 0.1967(133) & 0.0097(51) & 0.1974(136) & 0.0098(52)\\
\hline
1.030   & 0.1734(120) & 0.0095(46) & 0.1741(115) & 0.0095(45)\\
\hline
1.050   & 0.1567(112) & 0.0088(43) & 0.1574(112) & 0.0089(43)\\
\hline
1.070   & 0.1459(107) & 0.0083(41) & 0.1462(106) & 0.0083(41)\\
\hline
1.080   & 0.1413(101) & 0.0080(39) & 0.1418(103) & 0.0081(40)\\
\hline
1.090   & 0.1372(101) & 0.0079(35) & 0.1378(100) & 0.0078(33)\\
\hline
1.100   & 0.1341( 97) & 0.0076(37) & 0.1344( 95)   & 0.0077(35) \\
\hline
1.200   & 0.1115( 90) & 0.0075(35) & 0.1105( 83)   & 0.0063(32)\\
\hline 
1.300   & 0.0961( 70) & 0.0054(28) & 0.0965( 75 ) & 0.0055(27)\\
\hline
1.400   & 0.0860( 66) & 0.0048(25) & 0.0864( 65) & 0.0050(25)\\
\hline
 \end{tabular}
\end{center}
\end{table}

\begin{table}[ht!]
 \caption{Results for the lattice Coulomb potential, linear fits}
\begin{center}
\begin{tabular}{|l|c|c|c|r|}
\hline
    & \multicolumn{2}{c|}{$V=16\times10^{4}$}    & \multicolumn{2}{c|}{$V=16\times10^{3}$} \\
\hline
$\beta$ & $\alpha_{\mbox{\tiny layer}}$ &$\sigma_{\mbox{\tiny layer}}$ &$\alpha_{4D}$ &$\sigma_{4D}$ \\
\hline   
1.015   & 0.1806(122) & 0.0083(50) & 0.1812(124) & 0.0080(50)\\
\hline
1.030   & 0.1592(110) & 0.0082(45) & 0.1597(111) & 0.0080(50)\\
\hline
1.050   & 0.1438(103) & 0.0076(43) & 0.1444(103) & 0.0076(40)\\
\hline
1.070   & 0.1339( 98)  & 0.0072(40) & 0.1341( 98) & 0.0073(40)\\
\hline
1.080   & 0.1296( 93) & 0.0069(38) & 0.1301( 94) & 0.0070(40)\\
\hline
1.090   & 0.1259( 91) & 0.0067(36) & 0.1264( 92) & 0.0068(40)\\
\hline
1.100   & 0.1230( 89) & 0.0065(37) & 0.1234( 85) & 0.0064(35) \\
\hline
1.200   & 0.1024( 86) & 0.0076(36)  & 0.1013( 76)   & 0.0055(33)\\
\hline 
1.300   & 0.0883( 66) & 0.0047(27) & 0.0884( 69)  & 0.0051(30)\\
\hline
1.400   & 0.0790( 60) & 0.0042(25) & 0.0793( 60)  & 0.0040(20)\\
\hline
 \end{tabular}
\end{center}
\end{table}
\newpage

\begin{figure}[h!]
\centering
\subfloat[The $\sim \frac{1}{R}$ behavior of the potential in the layer phase 
and comparison with the 4D analog.]
{\includegraphics[width=5cm,angle=270]{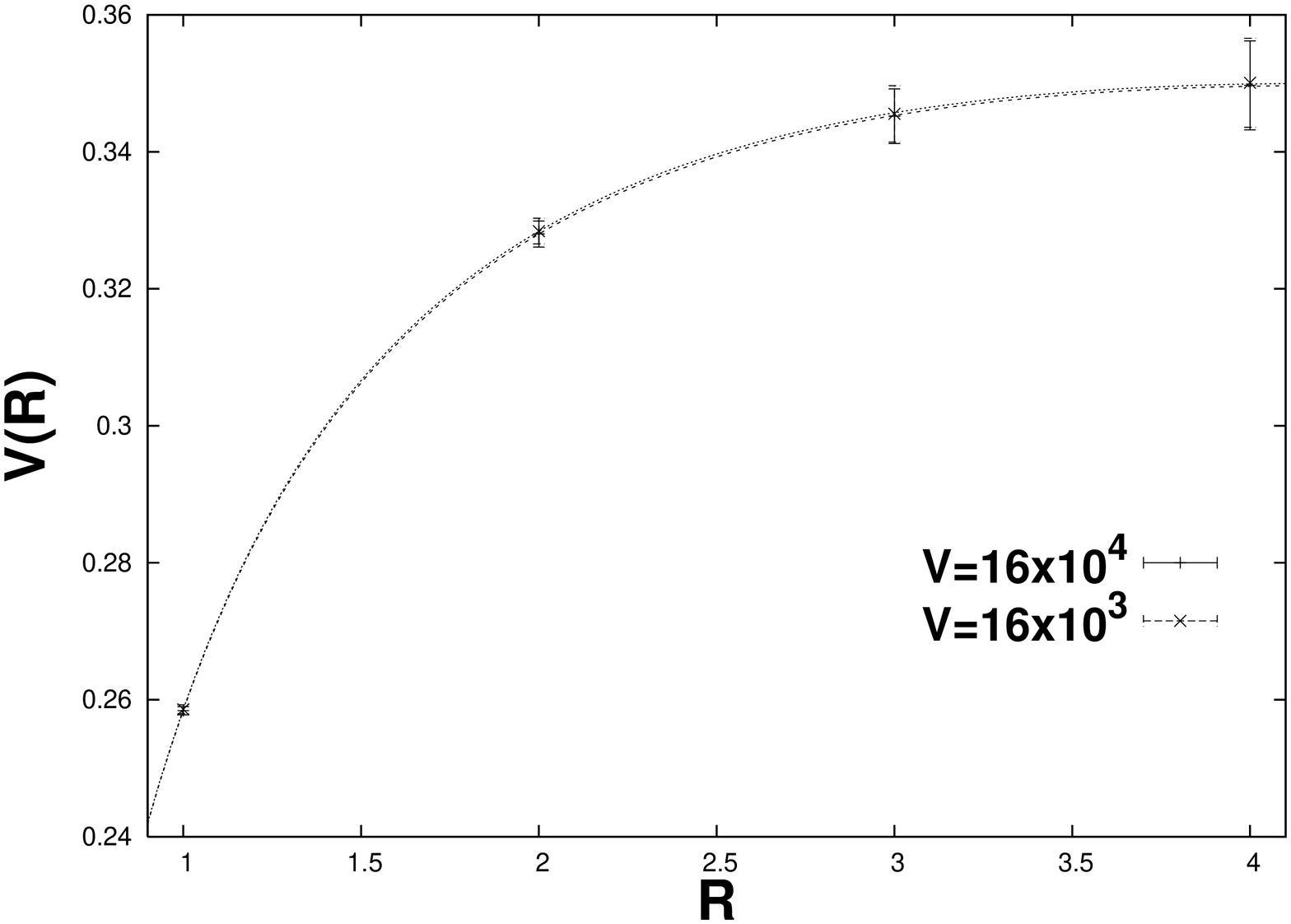}}
\subfloat[Results for $\alpha$ from 4D and 5D in the layer phase using linear fit
with T in the logarithm of the Wilson loops for the lattice potential.The lines 
come from the fitting with equation (2.16) and are identical.]
{\includegraphics[width=5cm,angle=270]{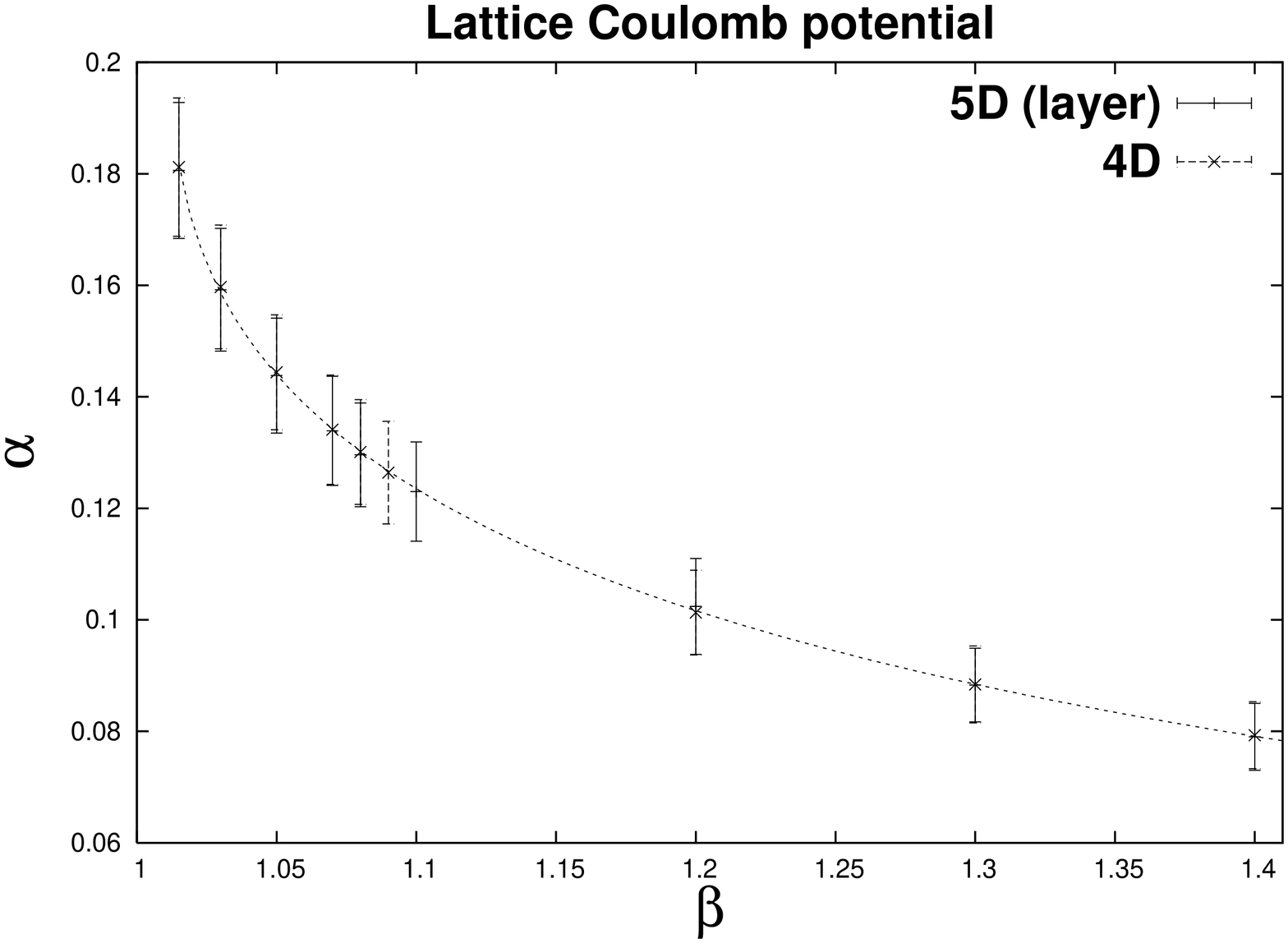}}\quad
\caption{}
\end{figure}

\subsubsection{\mbox{The $12^5$ case}}
The analysis presented above is now  repeated for a larger system, $12^{5}$ at an 
attempt to further strengthen our results. For the extraction of the potential from
Wilson loops only equation (2.12) has been used. Although
we had to restrict ourselves to smaller Wilson loops, due to the smaller extend 
of the lattice in the ''time'' direction, and utilize a much  
larger statistics for the 5D model the aforementioned picture does not change.

\begin{figure}[h!]
\centering
\subfloat[Comparison of the 4d layer potential with the usual 4D potential
for $\beta$=1.025 and lattice volumes $12^{5}$ and $12^{4}$.]
{\includegraphics[width=5cm,angle=270]{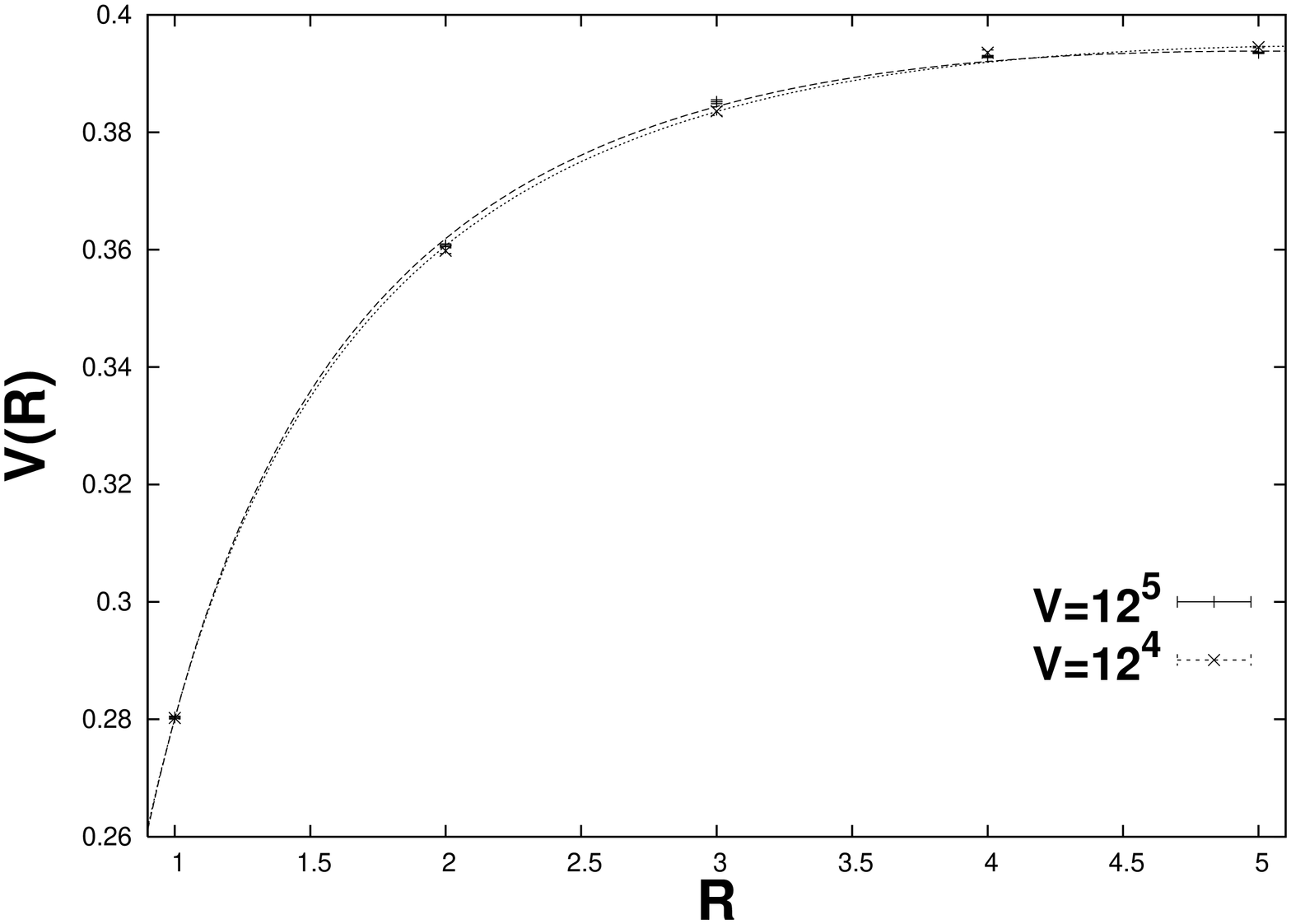}}\quad
\subfloat[Potentials from the 4d layer phase using the $12^{5}$ lattice volume.]
{\includegraphics[width=5cm,angle=270]{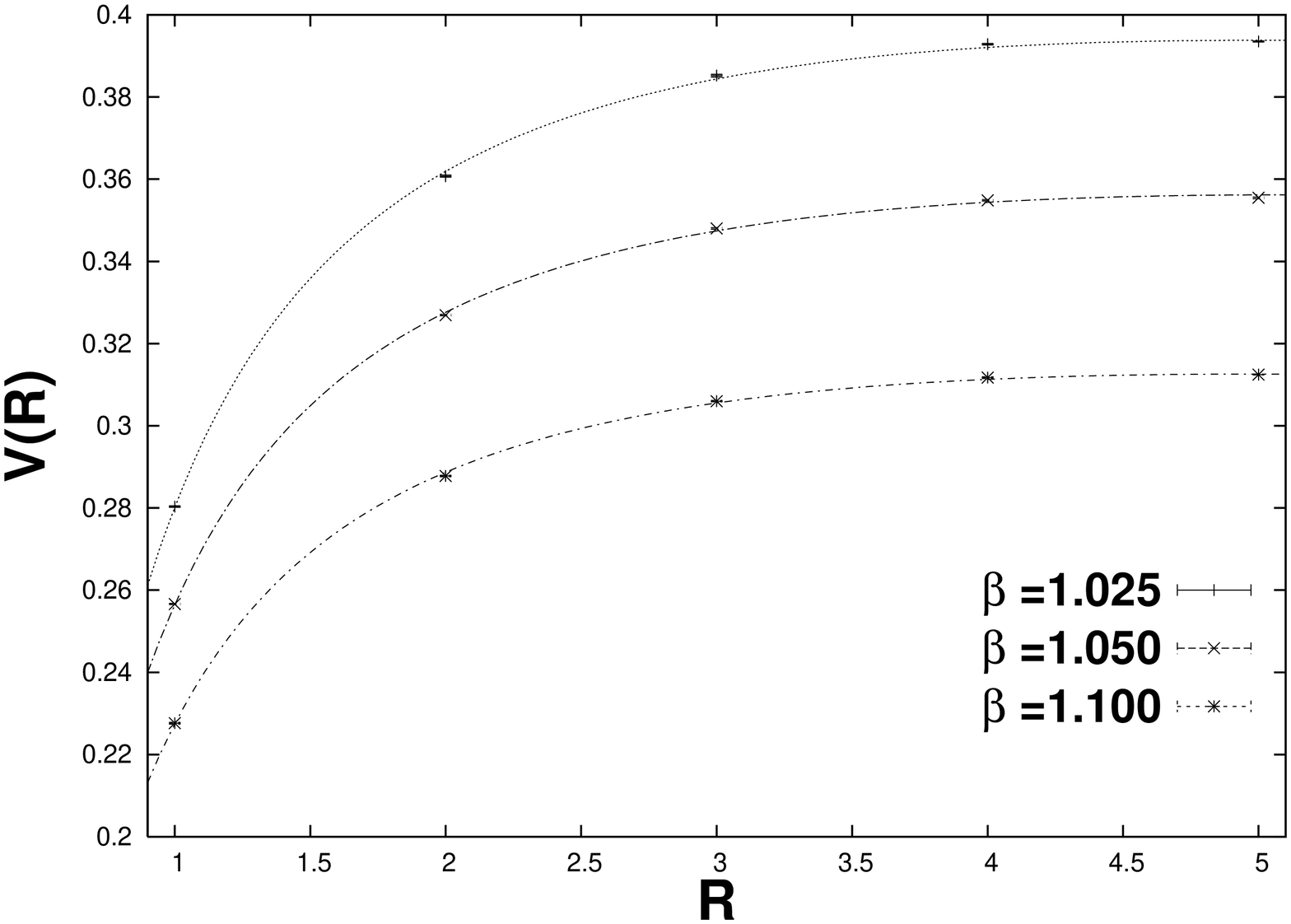}}
\caption{}
\end{figure}

We present our results for the potential V(R) in Figure 6 ((a) and (b)). The continuum
Coulomb potential ($\frac{1}{R}$) is used to fit the data from R=1 to R=5 with extremely
good accuracy, $\chi^{2}$ is always in the range 0.8-1.1.
\newpage
 \begin{table}[ht!]
 \caption{Results for the continuum Coulomb potential}
\begin{center}
\begin{tabular}{|l|c|c|c|r|}
\hline
    & \multicolumn{2}{c|}{$V=12^{5}$}       & \multicolumn{2}{c|}{$V=12^{4}$}   \\
\hline
$\beta$ & $\alpha_{\mbox{\tiny layer}}$ &$\sigma_{\mbox{\tiny layer}}$ &$\alpha_{4D}$ &$\sigma_{4D}$ \\
\hline   
1.015   & 0.1898(32) & 0.0080(20) & 0.1906(16) & 0.0075(15)\\
\hline
1.025   & 0.1778(40) & 0.0070(10) & 0.1730(15) & 0.0060(10)\\
\hline
1.050   & 0.1541(33) & 0.0060(10) & 0.1537(12) & 0.0060(10)\\
\hline
1.100   & 0.1333(26) & 0.0050( 5) &  0.1338(13) & 0.0050(10) \\
\hline
 \end{tabular}
\end{center}
\end{table}

 \begin{table}[ht!]
 \caption{Results for the lattice Coulomb potential}
\begin{center}
\begin{tabular}{|l|c|c|c|r|}
\hline
    & \multicolumn{2}{c|}{$V=12^{5}$}     & \multicolumn{2}{c|}{$V=12^{4}$}  \\
\hline
$\beta$ & $\alpha_{\mbox{\tiny layer}}$ &$\sigma_{\mbox{\tiny layer}}$ &$\alpha_{4D}$ &$\sigma_{4D}$ \\
\hline   
1.015   & 0.1735(10) & 0.0060(20) & 0.1708(14) & 0.0040(2)\\
\hline
1.025   & 0.1588(10) & 0.0052(20) & 0.1552(12) & 0.0040(2)\\
\hline
1.050   & 0.1386( 6) & 0.0045(14) & 0.1366(10) & 0.0038(2)\\
\hline
1.100   & 0.1184( 5) & 0.0037(11) &  0.1182(10) & 0.0035(2) \\
\hline
 \end{tabular}
\end{center}
\end{table}

\begin{figure}[!h]
\centering
\subfloat[All the results for $\alpha$ for the continuum Coulomb potential
from this section]
{\includegraphics[width=5cm,angle=270]{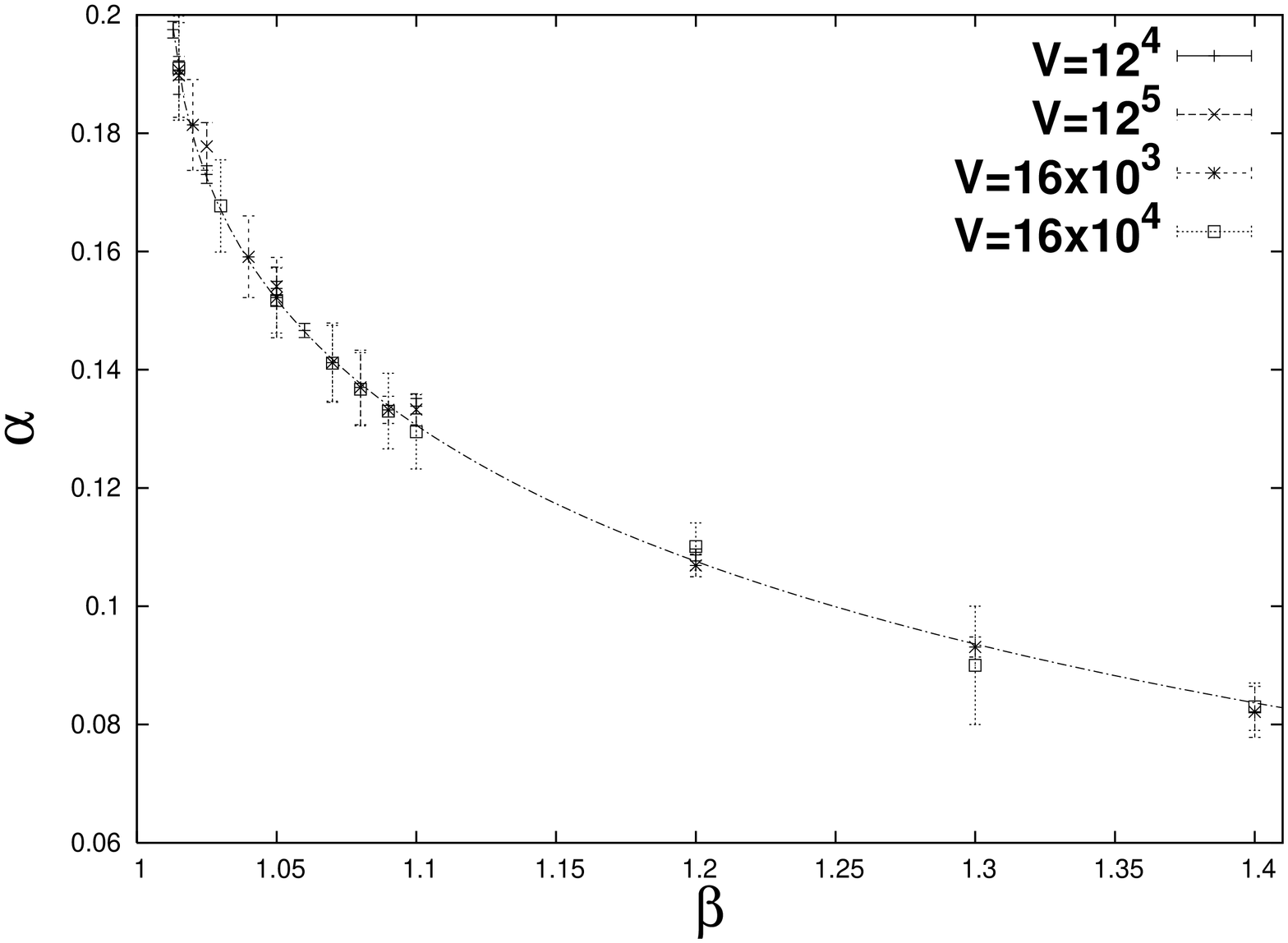}}\quad
\subfloat[All the results for $\alpha$ using the lattice Coulomb potential
from this section]
{\includegraphics[width=5cm,angle=270]{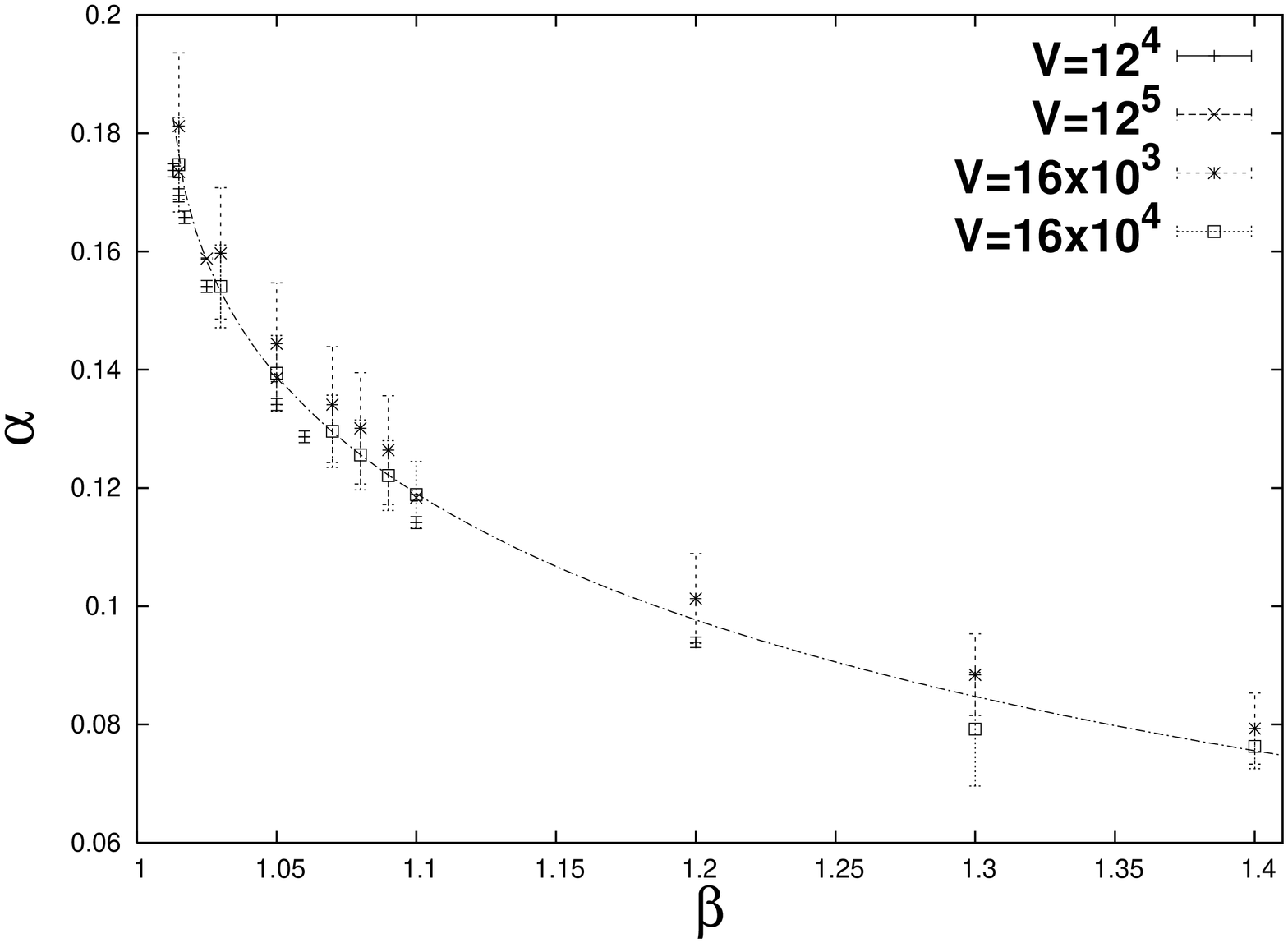}}
\caption{}
\end{figure}
As it is evident from Figure 6 and Tables 8 \& 9, where we compare results for the
two different systems: $12^{4}$ in the Coulomb phase and $12^{5}$ in the layer phase,
for the same $\beta$, the  presented consistency cannot pass unnoticed. It is not only the 
qualitative characteristics of the layer phase that point to the 4-dimensional 
nature of the forces governing the layers but also the quantitative agreement with
results from the pure 4D model. We found that all results from this section
for the effective fine structure constant fall on the same region of values making
them almost indistinguishable as Figure 7 shows.

\newpage
\subsubsection{The renormalized fine structure constant, a summary of results}
We find with the help of equation (2.16) and Tables 4 and 5, fitting $\alpha(\beta)$
 using $\beta_{c}=1.0111331(21)$ that:
\begin{flushleft}
$\alpha_{c-cc}$=0.230(30) \quad  $\lambda_{cc}$=0.32(10) \quad
$\alpha_{c-lc}$=0.210(30) \quad $\lambda_{lc}=0.32(10)$ \qquad \quad ($V=16\times10^{4}$) \quad 
 \\
 $\alpha_{c-cc}$=0.230(25) \quad  $\lambda_{cc}$=0.31( 8) \quad
$\alpha_{c-lc}$=0.208(24) \quad  $\lambda_{lc}$ = 0.33(10) \qquad \quad ($V=16\times10^{3}$) \quad  
\\
And for the linear fits, from Tables 6 and 7 we have:\\
$\alpha_{c-cc}$=0.235(49) \quad  $\lambda_{cc}$=0.32(16) \quad
$\alpha_{c-lc}$=0.216(45) \quad $\lambda_{lc}=0.32(16)$ \qquad \quad ($V=16\times10^{4}$)\\
$\alpha_{c-cc}$=0.238(52) \quad  $\lambda_{cc}$=0.31(16) \quad
$\alpha_{c-lc}$=0.219(48) \quad $\lambda_{lc}=0.31(16)$ \qquad \quad  ($V=16\times10^{3}$)\\
Finally from Tables 8 and 9 we have:\\
$\alpha_{c-cc}$=0.201(14) \quad $\lambda_{cc}$=0.51(22) \quad
$\alpha_{c-lc}$=0.198(08) \quad $\lambda_{lc}=0.395(60)$ \qquad \quad($V=12^{5}$)\\
$\alpha_{c-cc}$=0.209(07) \quad $\lambda_{cc}$=0.38(05) \quad
$\alpha_{c-lc}$=0.200(04) \quad $\lambda_{lc}=0.334(25)$ \qquad \quad($V=12^{4}$)\\
\end{flushleft}
using the continuum ($\bf{cc}$) and lattice Coulomb ($\bf{lc}$) potential respectively.
\footnote{Although the values of $\alpha_{c}$ come very close to
the value $\frac{\pi}{12}$ predicted at large R from the picture
of the rough string \cite{LSW}, this should be ascribed to the 
small four dimensional volume.}.

We can conclude that the layer-layer interactions are negligible and as a result
the interaction between two charges on a layer is a long range Coulomb interaction
with a massless carrier the photon.
\subsubsection{Results from the helicity modulus}
The main effort, as far as the h.m is concerned, was focused on volumes $12^{4}$ and
$12^{5}$ for the four and five dimensional systems respectively. We supplement the 
5D results with data from our previous work \cite{us}.

\begin{figure}[!hts]
\begin{center}
\includegraphics[width=7cm,angle=270]{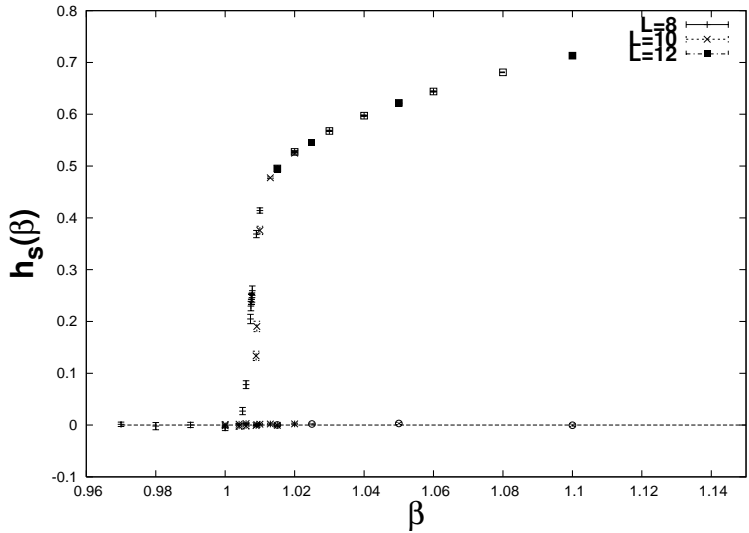}
\caption[]{The space and transverse helicity modulus (zero value points) as we perform the
transition from the 5D confining phase to the layer one, for various lattice volumes and $\beta^{'}$=0.20}
\end{center}
\label{f-3-3}
\end{figure}
\newpage
As the Figure 8 and Table 10 reveal, the 4d subspaces (layers) of our model realize the
above transition the exact same way as a 4D system realizes the passage from a confining
phase to the Coulomb phase. The transverse h.m ($h_{5}(\beta^{'})$) remains zero throughout
the transition, indicating confinement through the fifth direction while at the same time the
space h.m ($h_{S}(\beta)$), measured on the layers, obtains the same values as the corresponding
quantity of the four dimensional model.
 \begin{table}[ht!]
 \caption{Results for the helicity modulus and the corresponding values of $\alpha$}
\begin{center}
\begin{tabular}{|l|c|c|c|r|}
\hline
    & \multicolumn{2}{c|}{$V=12^{5}$}     & \multicolumn{2}{c|}{$V=12^{4}$}  \\
\hline
$\beta$ & $h(\beta)_{\mbox{\tiny layer}}$ &$\alpha_{\mbox{\tiny layer}}$ & $h(\beta)_{\mbox{\tiny 4D}}$ &$\alpha_{\mbox{\tiny 4D}}$\\
\hline   
1.015   & 0.4941(22) & 0.1611(7) & 0.4908(8) & 0.1622(4)\\
\hline
1.025   & 0.5455(15) & 0.1460(5) & 0.5462(5) & 0.1458(3)\\
\hline
1.050   & 0.6216(14) & 0.1281(4) & 0.6196(4) & 0.1285(2) \\
\hline
1.100   & 0.7134( 9)  & 0.1116(4) & 0.6196(5) & 0.1116(2) \\
\hline
1.200   & 0.8526( 6)  & 0.0934(3) & 0.8520(3) & 0.0935(1) \\
\hline
 \end{tabular}
\end{center}
\end{table}

Using the values we found from the helicity modulus (Table 10) and adopting the
behavior of equation (2.16) for $\alpha$  we have :\\
\newline
$\alpha_{\mbox{\tiny c-layer}}$=0.198(4) \quad $\lambda_{\mbox{\tiny layer}}$=0.28(2) \quad ($V=12^{5}$)\\ 
$\alpha_{c-4D}$=0.201(2) \quad $\lambda_{4D}$=0.276(8) \quad($V=12^{4}$)\\

These results are to be added to the ones of the previous subsection and the excellent
agreement for the renormalized fine structrure constant $\alpha_{c}$ with the lattice
Coulomb results must be noticed.

\section{The Coulomb phase}
Looking in the 5D ($\beta$,$\beta^{'}$) phase diagram (Figure 2) there is a separate
phase for big values of $\beta$ and $\beta^{'}$ which we mention as a Coulomb
5D phase. In order to characterize this phase we calculate the potential V(R) between two
heavy charges using the same techniques as in sections 2 and 3. 
If we follow the diagonal line $\beta$=$\beta^{'}$ there is a first order phase 
transition between the 5D strong phase and the 5D Coulomb phase for approximately
$\beta$=$\beta^{'}$=0.74 as we show in Figure 9.

For this part of our study we go deep in the 5D Coulomb phase following the diagonal line
of $\beta$=$\beta^{'}$ in the phase diagram for the biggest volume under study ($12^{5}$). By taking the gauge couplings $\beta,\beta^{'}$ equal, the previous anisotropy
of the model is now lost. As a consequence most equations used in the previous sections 
recive an almost natural generalization to five dimensions. 
\begin{figure}[h!]
\centering
\subfloat[Hysteresis loop of the mean value of the plaquette for a lattice volume
 $V=8^{5}$ ]
{\includegraphics[width=5cm,angle=270]{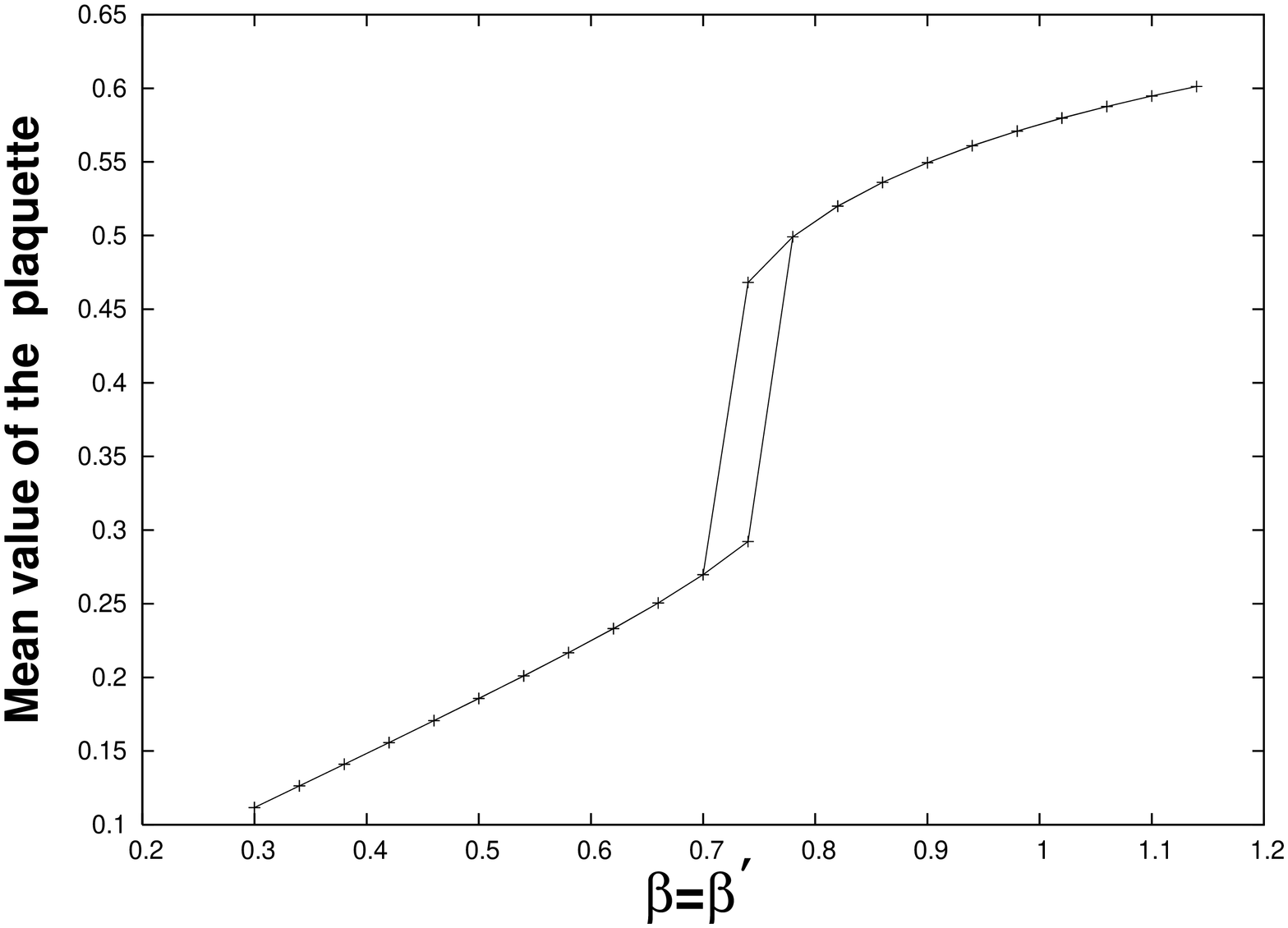}}
\subfloat[Hysteresis loop for the helicity modulus $h_{5D}(\beta)$, $V=8^{5}$.]
{\includegraphics[width=5cm,angle=270]{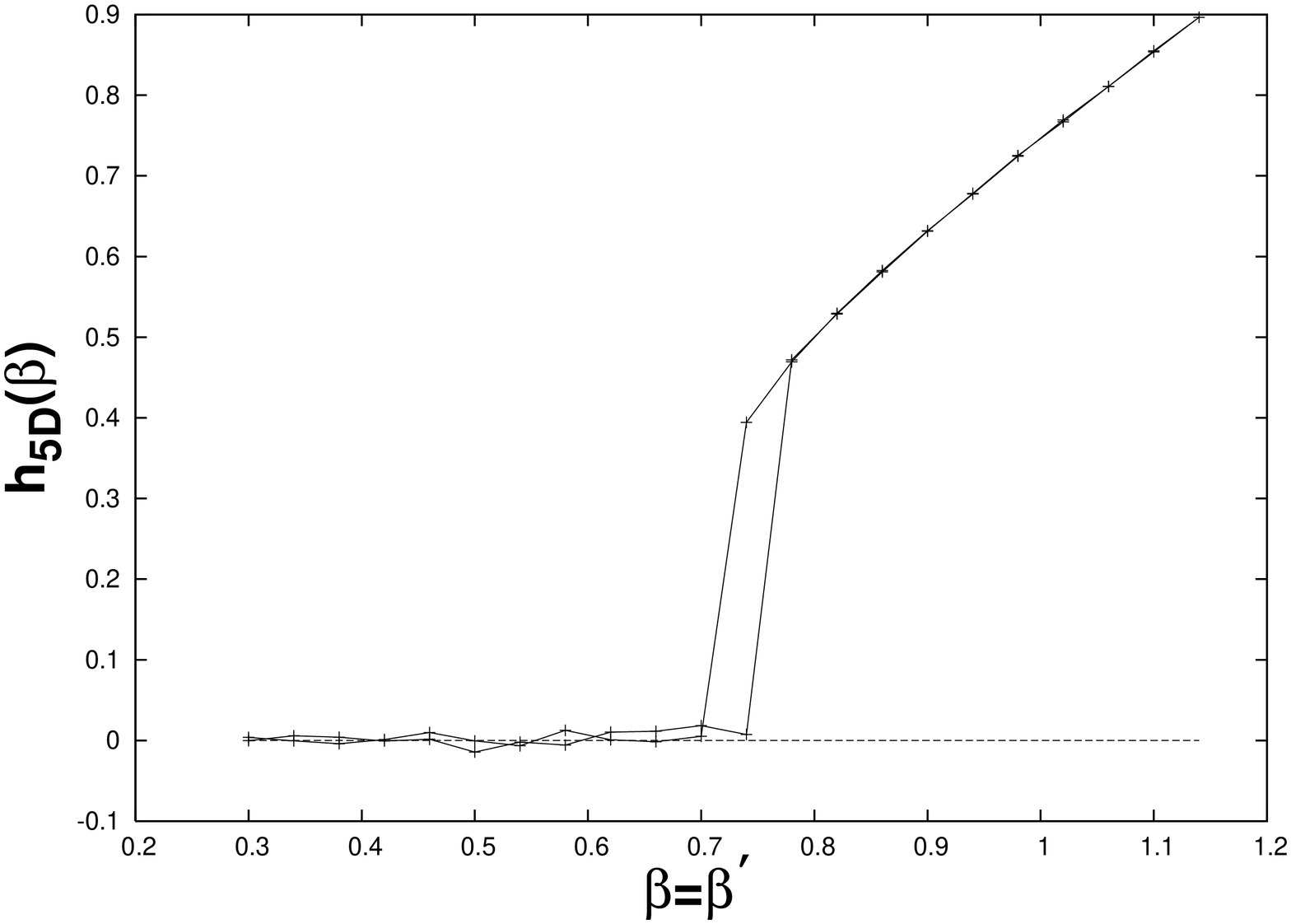}}\quad
\caption{}
\end{figure}
\newpage
Due to the passage from a 4-dimensional world (3+1) to a higher dimensional one
(n + 3 + 1 ) the form of the Coulomb potential changes. The $\sim \frac{1}{r}$
behavior no longer holds. The extra dimensions add powers to the denominator resulting to a
$\frac{1}{r^{1+n}}$ power law. We remind that:\\
\begin{displaymath}
 V(r)\varpropto \int { \frac{d^{3+n}k}{(2\pi)^{3+n}} \quad \mbox{\large e}^{i\vec{k}\vec{r}} \frac{1}{\vec{k}^{2}}} \quad \mbox{  and for the case n=1 we found  } 
 \frac{1}{4\pi^{2}r^{2}}
\end{displaymath}
using for the calculation spherical co-ordinates:
\begin{displaymath}
 \vec{k}=(k\sin\theta_{2}\sin\theta_{1}\cos\phi,k\sin\theta_{2}\sin\theta_{1}\sin\phi,
k\sin\theta_{2}\cos\theta_{1},k\cos\theta_{2})
\quad d^{4}k=k^{3}dkd\phi\sin\theta_{1}
d\theta_{1}\sin^{2}\theta_{2}d\theta_{2}
\end{displaymath}
\begin{displaymath}
\left ( 0<k<\infty,\quad 0<\phi<2\pi,\quad 0<\theta_{1}<\pi,\quad 0<\theta_{2}<\pi
\right ).
\end{displaymath}

Keeping all this in mind a natural generalization of equation (2.13) would be, for the 
case of a 5-dimensional Coulomb potential:\\
\begin{equation}
 V_{5D}(R)=const + \sigma_{5D}R + \frac{\hat{\alpha}_{5D}}{R^{2}}\quad
\mbox{with }\quad \hat{\alpha}_{5D}=\frac{e^{2}}{4\pi^{2}}\equiv\frac{\alpha_{5D}}{\pi}
\end{equation}
\newpage
\begin{figure}[!h]
\begin{center}
\includegraphics[width=8cm,angle=270]{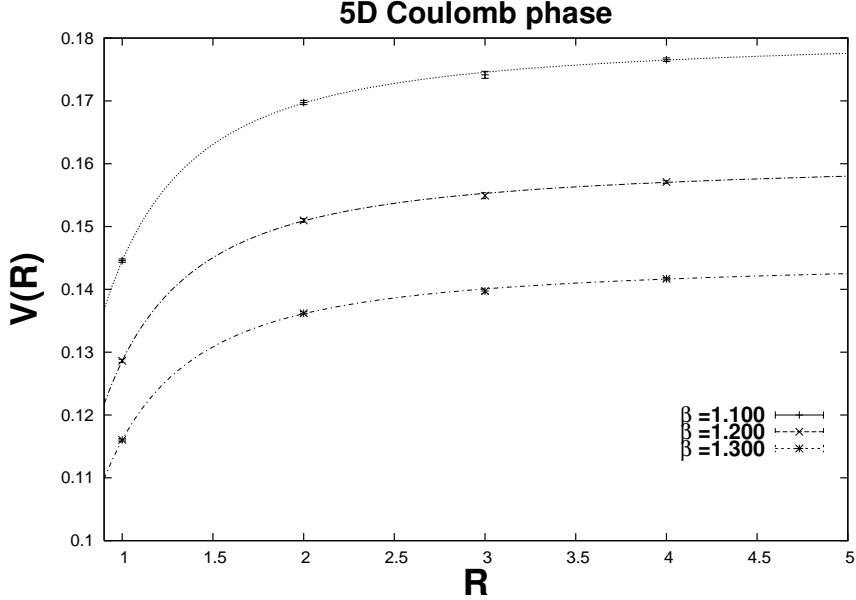}
\caption[]{Potential in the 5D Coulomb phase, Vol=$12^{5}$ for different values of
the coupling $\beta=\beta^{'}$. The error bars are covered by the used symbols.}
\end{center}
\label{f-4-1}
\end{figure}

 \begin{table}[ht]
 \caption{Results from the 5D Coulomb potentials and V=$12^{5}$}
\begin{center}
\begin{tabular}{|l|c|c|c|r|}
\hline
$\beta$    & $\hat{\alpha}_{5D}$ & $\sigma_{5D}$     & $\chi^{2}$/d.o.f  \\
\hline
1.100   & 0.0330(9) & 0.00033(24) & 0.67        \\
\hline
1.200   & 0.0293(8) & 0.00033(25) & 0.74         \\
\hline
1.300   & 0.0264(6) & 0.00028(18) & 0.66  \\
\hline
 \end{tabular}
\end{center}
\end{table}

Equation (4.1) describes well our data (Figure 10 and Table 11) while at
the same time additional endorsement comes from  the fact that all attempts
to use the 4-dimensional potentials for the description of the data were fruitless,
with a $\chi^{2}_{d.o.f}$ ranging from 7 to 20,
 thus excluding any connection with a 4D law. Another point worth
mentioning is that even if we subtract the confining term ($\sigma_{5D}R$) from
equation (4.1) we still get acceptable results ($\chi^{2}_{d.o.f}$ $\simeq$ 1-1.2)
with the resulting change in the values of $\hat{\alpha}_{5D}$ beeing within errors. 

With the form of potential established to $\sim \frac{1}{R^{2}}$ we continue with the
generalization of equation (2.15) to five dimensions.
\begin{equation}
 V_{lc}^{5D}(R)=\frac{4 \pi^{2}}{L_{s}^{4}}\sum_{\overrightarrow{k}\neq 0}\frac{\mbox{\large e}^{i\vec{k}\vec{R}}}{\sum_{j=1}^{4}2(1-\cos(k_{j}))}, \qquad k_{j}=0,\frac{2\pi}{L_{s}},\dots,\frac{2\pi(L_{s}-1)}{L_{s}}
\end{equation}
\newline
\begin{equation}
V_{5D}(R)=\sigma^{5D}_{lc} R -\hat{\alpha}^{5D}_{lc} V^{5D}_{lc}(R) +const
\end{equation}
\newpage
\begin{table}[ht]
 \caption{Results from the 5D lattice Coulomb potential and V=$12^{5}$}
\begin{center}
\begin{tabular}{|l|c|c|c|r|}
\hline
$\beta$    & $\hat{\alpha}^{5D}_{lc}$ & $\sigma^{5D}_{lc}$     & $\chi^{2}$/d.o.f  \\
\hline
1.100   & 0.0298( 4) & 0.0005(3) & 1.68        \\
\hline
1.200   & 0.0265( 6) & 0.0004(2) & 1.90         \\
\hline
1.300   & 0.0239(10) & 0.0004(2) & 1.80  \\
\hline
 \end{tabular}
\end{center}
\end{table}

The above values are not as good, in terms of the $\chi^{2}$, as the ones obtained
from the continuum Coulomb potential. They show a systematic deviation of order 10\%-12\%
from our previous results but that was also the case for the measurements of section 3.
Nevertheless they constitute a second estimate for the effective fine structure constant
in five dimensions.

Finally we go on and measure the helicity modulus for this phase to acquire our final 
estimate for $\alpha$. Due to the homogeneity of the model in the line $\beta=\beta^{'}$
 the choice of the plane in which the extra flux is imposed is not restricted,
since all the planes are now equivalent. Every possible choice will lead us to the same
result (a fact verified by our measurements). So, we continue with what we shall 
generally call helicity modulus in 5D ($h_{5D}(\beta)$) measured on the ($\mu \nu$ planes).  
 
\begin{equation}
 h_{5D}(\beta)=\frac{1}{(L_{\mu}L_{\nu})^{2}}\left( \left< \sum_{(\mu \nu)\mbox{\tiny planes}}(\beta\cos(\theta_{P}))\right>
-\left< (\sum_{(\mu \nu)\mbox{\tiny planes}}(\beta\sin(\theta_{P})))^{2} \right>\right)
\end{equation}

Lets pause here for a moment to consider the classical limit of the above equation. With
all fluctuations suppressed we have (following subsection 2.2) :\\
\begin{displaymath}
 S^{5D}_{\mbox{\tiny classical}(\Phi)}=\frac{1}{2}\beta\Phi^{2}\frac{V_{5D}}{(L_{\mu}L_{\nu})^{2}}=
\frac{1}{2}\beta\Phi^{2}\frac{L_{\mu}L_{\nu}L_{\rho}L_{\sigma}L_{\kappa}}{(L_{\mu}L_{\nu})^{2}}=
\frac{1}{2}\beta\Phi^{2}L_{\kappa}
\end{displaymath}
\begin{displaymath}
\rightarrow
F_{\mbox{\tiny classical}}(\Phi)-F_{\mbox{\tiny classical}}(0)=\frac{1}{2}\beta\Phi^{2}L_{\kappa}
\end{displaymath}
Again, with the replacement $\beta \rightarrow \beta_{R}$ and use of equation (2.10) we have
for the helicity modulus:
\begin{equation}
 h_{5D}(\beta)\sim \beta_{R}(\beta)L_{\kappa}
\end{equation}
Hence the h.m scales with the lattice length and as one approaches the infinite volume limit
the signal obtained from this quantity is infinitely enhanced. Although the argument presented above is based mainly on the classical approach, this is indeed the case and the helicity modulus
applied for the five dimensional system behaves exactly as equation (4.5) predicts. So, in order to extract the value of $\beta_{R}$, the appropriate rescaling is needed. To that end,
 all measurements in this section concerning the h.m are the product of the simple rescaling that equation (4.5) suggests $\left(h_{5D}(\beta)\rightarrow\frac{h_{5D}(\beta)}{L_{\kappa}}\right)$.
In Table 13 one can find the verification of all this where the estimates from the two lattice volumes ($8^{5}$ and $12^{5}$) are identical, within the error bars. 
  
\newpage
\begin{table}[ht!]
 \caption{The helicity modulus and the resulting values of $\hat{\alpha}$
for the five dimensional Coulomb phase and two lattice volumes $8^{5}$ and $12^{5}$}
\begin{center}
\begin{tabular}{|l|c|c|c|r|}
\hline
 &   \multicolumn{2}{c|}{$V=12^{5}$} & \multicolumn{2}{c|}{$V=8^{5}$}       \\
\hline
$\beta=\beta^{'}$  & $h(\beta)_{\mbox{\tiny 5D}}$ &$\hat{\alpha}_{\mbox{\tiny 5D}}$ &  $h(\beta)_{\mbox{\tiny 5D}}$ &$\hat{\alpha}_{\mbox{\tiny 5D}}$  \\
\hline   
0.800  & 0.5017(4) & 0.0505(2) & 0.5014(4) & 0.0506(2) \\
\hline
0.900   & 0.6312(4) & 0.0402(2) & 0.6306(5) & 0.0402(3)\\
\hline
1.000   & 0.7460(3) & 0.0339(2) & 0.7458(3) & 0.0340(2)\\
\hline
1.100   & 0.8547(3)  & 0.0297(1) & 0.8539(7) & 0.0297(1)\\
\hline
1.200   & 0.9611(2)  & 0.0264(1) & 0.9610(2) & 0.0264(1) \\
\hline
1.300   & 1.0653(3)  & 0.0238(1) & 1.0657(2) & 0.0238(1) \\
\hline
1.400   & 1.1693(2)  & 0.0217(1) & 1.1694(2) & 0.0217(1) \\
\hline
 \end{tabular}
\end{center}
\end{table}
As Table 13 shows the agreement between the values of $\hat{\alpha}$ as they are obtained
from the five dimensional helicity modulus and the corresponding lattice Coulomb 
potential (Table 12) is almost perfect. The measurements of the helicity modulus 
are extended near the critical point in order to sketch the behavior of the effective
renormalized charge $\hat{\alpha}_{5D}(\beta)$ as we approach the transition. We observe
that there is no volume dependence as the difference between the results from the two 
lattice volumes is within the statistical error. 
\section{Conclusions}
Throughout this whole investigation we observed  no discrepancy between the
two systems at any point: The 4D pure U(1) gauge model in the Coulomb phase
and the anisotropic 5D U(1) model in the layer phase exhibit exactly the same 
behavior. From the values and the form of the potential, to the estimates for the 
renormalized coupling and the string tension and finally to the values of 
$\alpha_{c}$ and $\lambda$. All  signals point to the four dimensional
nature of the long range interactions in the layers and the presence of a 
massless particle, the photon, albeit the need for larger volumes.
The obtained agreement respponds, indirectly, on another subtile matter, 
which is the 
role of the layer-layer interactions in the physical picture.
It seems, to the extend that we can observe, that there is no evidence of any 
significant influence that could lead to an essential alteration from the 
known four dimensional laws.   
To clarify this point we would like to add a few remarks regarding the nature
of the gauge particle.
Fu and Nielsen, in a followup
work \cite{FuNi}, have thoroughly examined the properties of this gauge particle. 
Their analysis suggested, from a strong coupling expansion point of view, that
to lowest-order approximation the photon propagator is identical with the one in the 
isotropic 4-dimensional U(1) gauge field model. But, upon corrections the propagator
received contributions from the layer-layer coupling ($\beta^{'}$). Taking into account
the contributions of all graphs consisting of plaquettes connecting neighboring layers
(by means of an effective action) they have managed to show that to order $\beta^{'4}$ 
\begin{displaymath}
 \mbox{photon propagator}= \mbox{ordinary photon propagator}\times \left(1-\frac{1}{48}\frac{\beta^{'4}}{\beta}\right)
\end{displaymath}
for links in the same layer.
\footnote{Ordinary propagator means a propagator for which $\beta^{'}$=0 so that all
layers are isolated from each other.}
But, for the range of our measurements, corrections of this order of magnitude are
put into shadow by our error estimates. It is  beyond our measuring capabilities to
examine the  possible reprecautions of such a term and the differentiations that it
could lead. On the optimistic side though, for the whole range of the parameters that
we used for our analysis this ``correction'' ranges from 0.999967 (starting point)
to 0.999976 (end point). So we strongly doubt that any meaningfull diversion from
the four dimensional law can be found.

A second point that deserves attention is the use of the helicity modulus for
the extraction of the renormalized coupling ($\beta_{R}(\beta)$). The obtained
information from the use of this quantity is in very good agreement with the
results obtained from the traditional method of determination of $\beta_{R}$
using the Wilson loops,
with one advantage: it is much cheaper, from a computer power(/time) point of
view, to use the helicity modulus than to resort to Wilson loops. The 
required information comes directly from a single measurement, without any 
intermediate steps, thus reducing drastically the complexity of the method, 
compared to the usual approach. To that end (and) for the characterization of the
various phases of our model we find the helicity modulus a much better tool 
than Wilson loops. 

Finally, in the 5D Coulomb phase we found that  
the values of the effective $\alpha_{5D}$(=$\pi\hat{\alpha}_{5D}$) are smaller than
the values of effective $\alpha_{4D}$=$\alpha_{\mbox{\tiny layer}}$ as our results indicate
and also that the effective $\alpha_{5D}$ is slightly bigger than the bare coupling $\alpha_{0}$=$\frac{g_{0}^{2}}{4\pi}$=$\frac{1}{\beta}\frac{1}{4\pi}$
, where $g_{0}^{2}$=$\frac{g_{5}^{2}}{\alpha}$ 
$\simeq$ $g_{5}^{2} \Lambda_{UV}$ the dimensionless 5D bare U(1) gauge coupling
\footnote{here $\alpha$ is the 5D lattice spacing and $\Lambda_{UV}$ the ultraviolet cutoff.}.
The values of $\beta$ that we used for the determination of the 5D Coulomb potential
are far away from the phase transition. In order
for someone to be able to compare results between 4D and 5D an extrapolation up
to the critical point is neccesary. To that end, we calculated the helicity moduli
for two different volumes (Table 13) and fitted the results using equation (2.16) 
under the assumption that there is no drastic change in the behavior of $\hat{\alpha}$
as we go to five dimensions and consequently, that the equation remains still valid 
 for the particular case under study. We used the critical value of $\beta$ as a
free parameter and found: (i) that there is no volume dependence and (ii) the
renormalized fine structure constant takes the value of 
$\alpha_{c-5D}$=0.218(48), a value near to the one in four dimensions and 
$\lambda$=0.49(37). Unfortunately
the quality of the fit was not  pleasing ($\chi^{2}\mbox{\tiny d.o.f}=0.0007$, hence the
large errors) but it did manage to reproduce the value of the critical $\beta$ in the correct region ($\beta_{c}=0.741$).

Because the 5D QED is not a pertubativelly
renormalizable field theory and the gauge coupling has Mass dimensions ($g_{5}^{2}\sim M^{-1}$)
, powers of the cutoff $\Lambda_{UV}$ appear in the calculations of loops corrections 
(see \cite{Dien} and references therein) to the vertex and self energy graphs. In the
lattice calculations the cutoff does not appear explicitly but only implicitly through
the volume dependence of  Monte Carlo results. So, in the presence of matter fields,
we expect a strong volume dependence
for the effective renormalized charge $\alpha_{5D}(\beta)$ in the extrapolation to the 
critical $\beta$ and to the infinite volume, different from what we found in the present 
paper for the pure U(1). This study is outside the scope of this work but we believe 
that it deserves further investigation.

\section{Acknowledgements}
We acknowledge support from the EPEAEK programme ``Pythagoras II'' co-funded by the
European Union (75\%) and the Hellenic State (25\%). We are very greatfull to 
P.de Forcrand for a series of very usefull conversations regarding the helicity 
modulus and its applications and to G.Koutsoumbas, K.Anagnostopoulos and P.Dimopoulos
for their help and support, for reading and discusing this manuscript.    


 



\end{document}

%% file: macros.tex
\newcommand{\ba}{\begin{array}}
\newcommand{\ea}{\end{array}}

\newcommand{\Dslash}{\relax{\kern+.25em / \kern-.70em D}}

\newcommand{\Real}{\relax{\mathsf{\Gamma\kern-.35em R}}}
\newcommand{\Int}{\relax{\mathsf{Z\kern-.40em Z}}}

\newcommand{\gbar}{\kern1pt\overline{\kern-1pt g\kern-0pt}\kern1pt}
\newcommand{\mbar}{\kern2pt\overline{\kern-1pt m\kern-1pt}\kern1pt}
\newcommand{\obar}[1]{\kern3pt\overline{\kern-2pt #1\kern-0pt}\kern1pt}



%% file: PDG.tex
\setlength{\unitlength}{0.240900pt}
\ifx\plotpoint\undefined\newsavebox{\plotpoint}\fi
\sbox{\plotpoint}{\rule[-0.200pt]{0.400pt}{0.400pt}}%
\begin{picture}(1500,900)(0,0)
\sbox{\plotpoint}{\rule[-0.200pt]{0.400pt}{0.400pt}}%
\put(181.0,123.0){\rule[-0.200pt]{4.818pt}{0.400pt}}
\put(161,123){\makebox(0,0)[r]{ 0.2}}
\put(1419.0,123.0){\rule[-0.200pt]{4.818pt}{0.400pt}}
\put(181.0,205.0){\rule[-0.200pt]{4.818pt}{0.400pt}}
\put(161,205){\makebox(0,0)[r]{ 0.4}}
\put(1419.0,205.0){\rule[-0.200pt]{4.818pt}{0.400pt}}
\put(181.0,287.0){\rule[-0.200pt]{4.818pt}{0.400pt}}
\put(161,287){\makebox(0,0)[r]{ 0.6}}
\put(1419.0,287.0){\rule[-0.200pt]{4.818pt}{0.400pt}}
\put(181.0,369.0){\rule[-0.200pt]{4.818pt}{0.400pt}}
\put(161,369){\makebox(0,0)[r]{ 0.8}}
\put(1419.0,369.0){\rule[-0.200pt]{4.818pt}{0.400pt}}
\put(181.0,451.0){\rule[-0.200pt]{4.818pt}{0.400pt}}
\put(161,451){\makebox(0,0)[r]{ 1}}
\put(1419.0,451.0){\rule[-0.200pt]{4.818pt}{0.400pt}}
\put(181.0,532.0){\rule[-0.200pt]{4.818pt}{0.400pt}}
\put(161,532){\makebox(0,0)[r]{ 1.2}}
\put(1419.0,532.0){\rule[-0.200pt]{4.818pt}{0.400pt}}
\put(181.0,614.0){\rule[-0.200pt]{4.818pt}{0.400pt}}
\put(161,614){\makebox(0,0)[r]{ 1.4}}
\put(1419.0,614.0){\rule[-0.200pt]{4.818pt}{0.400pt}}
\put(181.0,696.0){\rule[-0.200pt]{4.818pt}{0.400pt}}
\put(161,696){\makebox(0,0)[r]{ 1.6}}
\put(1419.0,696.0){\rule[-0.200pt]{4.818pt}{0.400pt}}
\put(181.0,778.0){\rule[-0.200pt]{4.818pt}{0.400pt}}
\put(161,778){\makebox(0,0)[r]{ 1.8}}
\put(1419.0,778.0){\rule[-0.200pt]{4.818pt}{0.400pt}}
\put(181.0,860.0){\rule[-0.200pt]{4.818pt}{0.400pt}}
\put(161,860){\makebox(0,0)[r]{ 2}}
\put(1419.0,860.0){\rule[-0.200pt]{4.818pt}{0.400pt}}
\put(181.0,123.0){\rule[-0.200pt]{0.400pt}{4.818pt}}
\put(181,82){\makebox(0,0){ 0}}
\put(181.0,840.0){\rule[-0.200pt]{0.400pt}{4.818pt}}
\put(321.0,123.0){\rule[-0.200pt]{0.400pt}{4.818pt}}
\put(321,82){\makebox(0,0){ 0.2}}
\put(321.0,840.0){\rule[-0.200pt]{0.400pt}{4.818pt}}
\put(461.0,123.0){\rule[-0.200pt]{0.400pt}{4.818pt}}
\put(461,82){\makebox(0,0){ 0.4}}
\put(461.0,840.0){\rule[-0.200pt]{0.400pt}{4.818pt}}
\put(600.0,123.0){\rule[-0.200pt]{0.400pt}{4.818pt}}
\put(600,82){\makebox(0,0){ 0.6}}
\put(600.0,840.0){\rule[-0.200pt]{0.400pt}{4.818pt}}
\put(740.0,123.0){\rule[-0.200pt]{0.400pt}{4.818pt}}
\put(740,82){\makebox(0,0){ 0.8}}
\put(740.0,840.0){\rule[-0.200pt]{0.400pt}{4.818pt}}
\put(880.0,123.0){\rule[-0.200pt]{0.400pt}{4.818pt}}
\put(880,82){\makebox(0,0){ 1}}
\put(880.0,840.0){\rule[-0.200pt]{0.400pt}{4.818pt}}
\put(1020.0,123.0){\rule[-0.200pt]{0.400pt}{4.818pt}}
\put(1020,82){\makebox(0,0){ 1.2}}
\put(1020.0,840.0){\rule[-0.200pt]{0.400pt}{4.818pt}}
\put(1159.0,123.0){\rule[-0.200pt]{0.400pt}{4.818pt}}
\put(1159,82){\makebox(0,0){ 1.4}}
\put(1159.0,840.0){\rule[-0.200pt]{0.400pt}{4.818pt}}
\put(1299.0,123.0){\rule[-0.200pt]{0.400pt}{4.818pt}}
\put(1299,82){\makebox(0,0){ 1.6}}
\put(1299.0,840.0){\rule[-0.200pt]{0.400pt}{4.818pt}}
\put(1439.0,123.0){\rule[-0.200pt]{0.400pt}{4.818pt}}
\put(1439,82){\makebox(0,0){ 1.8}}
\put(1439.0,840.0){\rule[-0.200pt]{0.400pt}{4.818pt}}
\put(181.0,123.0){\rule[-0.200pt]{303.052pt}{0.400pt}}
\put(1439.0,123.0){\rule[-0.200pt]{0.400pt}{177.543pt}}
\put(181.0,860.0){\rule[-0.200pt]{303.052pt}{0.400pt}}
\put(181.0,123.0){\rule[-0.200pt]{0.400pt}{177.543pt}}
\put(40,491){\makebox(0,0){$\beta$}}
\put(810,21){\makebox(0,0){$\beta^{'}$}}
\put(321,614){\makebox(0,0)[l]{$\bf{L}$}}
\put(461,205){\makebox(0,0)[l]{$\bf{S}$}}
\put(1020,614){\makebox(0,0)[l]{$\bf{C}$}}
\put(188,451){\raisebox{-.8pt}{\makebox(0,0){$\Diamond$}}}
\put(251,451){\raisebox{-.8pt}{\makebox(0,0){$\Diamond$}}}
\put(286,451){\raisebox{-.8pt}{\makebox(0,0){$\Diamond$}}}
\put(356,451){\raisebox{-.8pt}{\makebox(0,0){$\Diamond$}}}
\put(391,451){\raisebox{-.8pt}{\makebox(0,0){$\Diamond$}}}
\put(426,451){\raisebox{-.8pt}{\makebox(0,0){$\Diamond$}}}
\put(426,451){\raisebox{-.8pt}{\makebox(0,0){$\Diamond$}}}
\put(429,492){\raisebox{-.8pt}{\makebox(0,0){$\Diamond$}}}
\put(429,512){\raisebox{-.8pt}{\makebox(0,0){$\Diamond$}}}
\put(429,532){\raisebox{-.8pt}{\makebox(0,0){$\Diamond$}}}
\put(426,573){\raisebox{-.8pt}{\makebox(0,0){$\Diamond$}}}
\put(426,635){\raisebox{-.8pt}{\makebox(0,0){$\Diamond$}}}
\put(426,676){\raisebox{-.8pt}{\makebox(0,0){$\Diamond$}}}
\put(426,717){\raisebox{-.8pt}{\makebox(0,0){$\Diamond$}}}
\put(426,758){\raisebox{-.8pt}{\makebox(0,0){$\Diamond$}}}
\put(422,778){\raisebox{-.8pt}{\makebox(0,0){$\Diamond$}}}
\put(422,819){\raisebox{-.8pt}{\makebox(0,0){$\Diamond$}}}
\put(461,451){\raisebox{-.8pt}{\makebox(0,0){$\Diamond$}}}
\put(466,442){\raisebox{-.8pt}{\makebox(0,0){$\Diamond$}}}
\put(530,422){\raisebox{-.8pt}{\makebox(0,0){$\Diamond$}}}
\put(600,397){\raisebox{-.8pt}{\makebox(0,0){$\Diamond$}}}
\put(670,369){\raisebox{-.8pt}{\makebox(0,0){$\Diamond$}}}
\put(740,342){\raisebox{-.8pt}{\makebox(0,0){$\Diamond$}}}
\put(810,319){\raisebox{-.8pt}{\makebox(0,0){$\Diamond$}}}
\put(880,297){\raisebox{-.8pt}{\makebox(0,0){$\Diamond$}}}
\put(950,270){\raisebox{-.8pt}{\makebox(0,0){$\Diamond$}}}
\put(1020,246){\raisebox{-.8pt}{\makebox(0,0){$\Diamond$}}}
\put(1090,225){\raisebox{-.8pt}{\makebox(0,0){$\Diamond$}}}
\put(1194,205){\raisebox{-.8pt}{\makebox(0,0){$\Diamond$}}}
\put(1299,191){\raisebox{-.8pt}{\makebox(0,0){$\Diamond$}}}
\put(1439,184){\raisebox{-.8pt}{\makebox(0,0){$\Diamond$}}}
\put(181.0,123.0){\rule[-0.200pt]{303.052pt}{0.400pt}}
\put(1439.0,123.0){\rule[-0.200pt]{0.400pt}{177.543pt}}
\put(181.0,860.0){\rule[-0.200pt]{303.052pt}{0.400pt}}
\put(181.0,123.0){\rule[-0.200pt]{0.400pt}{177.543pt}}
\end{picture}